%% file: final_main.tex
\documentclass[manuscript,screen]{acmart}
\usepackage{booktabs,tabularx,array,colortbl,xcolor}
\definecolor{RowBlue}{RGB}{242,247,255}
\usepackage[table]{xcolor}
\usepackage{tabularx}
\usepackage{paralist}
\usepackage{bbm}

\usepackage[most]{tcolorbox}
\newtcolorbox{runningexamplebox}[1][]{%
  enhanced, breakable,
  colback=black!2,          % very light gray
  colframe=black!35,        % subtle gray border
  boxrule=0.6pt,
  arc=1.5mm,
  left=2mm,right=2mm,top=1mm,bottom=1mm,
  fonttitle=\bfseries,
  coltitle=black,
  title={Running example used throughout (TalkPlayData-2)},
  #1
}

\usepackage{booktabs}
\usepackage{enumitem}
\usepackage[dvipsnames]{xcolor}

\usepackage{booktabs}
\usepackage{amsmath} 
\usepackage{tabularx} \usepackage{changepage}
\usepackage[table]{xcolor}
\usepackage{tabularx}
\newcolumntype{L}[1]{>{\raggedright\arraybackslash}p{#1}}
\renewcommand{\arraystretch}{1.25}
\usepackage{subcaption}
\usepackage{amsmath} \usepackage{booktabs} \usepackage{tabularx}
\usepackage{xltabular}

\usepackage[most]{tcolorbox}

\tcbuselibrary{breakable}

\newcommand{\highlightbox}[1]{%
\begin{tcolorbox}[
  breakable,
  colback=dzrorange!15,
  colframe=dzrorange!15,
  boxrule=0pt,
  arc=0pt,
  left=5pt,
  right=5pt,
  top=5pt,
  bottom=5pt,
  boxsep=3pt,
  before skip=6pt,
  after skip=6pt
]
#1
\end{tcolorbox}%
}

\AtBeginDocument{%
  }

\setcopyright{acmlicensed}
\copyrightyear{2025}
\acmYear{2025}
\acmDOI{XXXXXXX.XXXXXXX}
\acmISBN{978-1-4503-XXXX-X/2018/06}

\begin{document}

\title{Music Recommendation with Large Language Models: \\Challenges, Opportunities, and Evaluation} %An Evaluation Agenda}

%\title{Challenges and Visions on the Use of Large Language Models in Music Recommendation \textcolor{blue}{Environments?}}

 \author{Elena V. Epure}
 \authornote{Equal contribution.}
\email{research@deezer.com}
\orcid{https://orcid.org/0000-0002-6930-9482}
\affiliation{
   \institution{Deezer Research}
  \city{Paris}
   \country{France}
   \and 
   \institution{Idiap Research Institute}
  \city{Martigny}
   \country{Switzerland}
 }

 \author{Yashar Deldjoo}
 \authornotemark[1]
\email{deldjooy@acm.org}
\orcid{https://orcid.org/0000-0002-6767-358X}
 \affiliation{
   \institution{ Politecnico di Bari}
   \city{Bari}
   \country{Italy}}

 \author{Bruno Sguerra}
\authornotemark[1]
 \email{research@deezer.com}
 \orcid{https://orcid.org/0000-0003-1158-9095}
 
 \affiliation{
   \institution{Deezer Research}
  \city{Paris}
   \country{France}
 }
 \author{Markus Schedl}
 \email{markus.schedl@jku.at}
 \orcid{https://orcid.org/0000-0003-1706-3406}
 \affiliation{%
  \institution{Johannes Kepler University Linz and Linz Institute of Technology}
   \city{Linz}
   \country{Austria}
 }

  \author{Manuel Moussallam}
 \email{research@deezer.com}
 \orcid{https://orcid.org/0000-0003-0886-5423}
 
 \affiliation{
   \institution{Deezer Research}
  \city{Paris}
   \country{France}
 }

% \affiliation{%

% \affiliation{%

% \affiliation{%

% \affiliation{%

\renewcommand{\shortauthors}{Epure et al.}
\renewcommand{\shorttitle}{Music Recommendation with Large Language Models}

\begin{abstract}
Music Recommender Systems (MRSs) have long relied on an information retrieval framing, where progress is measured mainly through accuracy on retrieval-oriented subtasks. 
While effective, this reductionist paradigm struggles to address the deeper question of what makes a good recommendation. Attempts to broaden evaluation, through user studies or fairness analyses, have had limited impact.
The emergence of Large Language Models (LLMs) disrupts this framework: LLMs are generative rather than ranking-based, making standard accuracy metrics questionable. They also introduce challenges such as hallucinations, knowledge cutoffs, non-determinism, and opaque training data, rendering traditional train/test protocols difficult to interpret. At the same time, LLMs create new opportunities, enabling natural language (NL) interaction and even allowing models to act as evaluators.

This work argues that the shift toward LLM-driven MRSs requires rethinking evaluation. We first review how LLMs have impacted user modeling, item modeling, and NL-based recommendation in music. We then analyze evaluation practices from NLP, highlighting methodologies and open challenges relevant to MRSs. Finally, we synthesize insights, focusing on how LLM prompting applies to MRSs, to outline a structured set of success and risk dimensions. Our goal is to provide the MRSs community with an updated, pedagogical, and cross-disciplinary perspective on evaluation.
\end{abstract}




%\received{November 2025}

\maketitle

\definecolor{dzrblue}{RGB}{150, 249, 243}
\definecolor{dzrorange}{RGB}{255, 218, 198}

\section{Introduction}
\label{sec:introduction}

At its origins, the field of Music Recommender Systems (MRSs) was built upon the same foundations as the broader Recommender Systems (RSs) domain, namely, by framing its objectives as a set of information retrieval tasks.
Practitioners adopting this view benefited from the rapid advancements in the RSs community and in machine learning in general.
For decades, progress in MRSs essentially meant improving performance on a series of retrieval-oriented subtasks, often evaluated offline, such as matrix completion, clustering, or sequence continuation.
Soon enough, this framing revealed its limitations due to its markedly reductionist nature.
It simplified the original question,~\textit{how do we design systems that make good recommendations?}, by sidestepping the harder one:~\textit{what actually constitutes a good recommendation?}
Instead, the focus shifted to a more tractable problem:~\textit{how can we build systems that accurately predict what users will eventually consume?}
While the two are arguably not the same, the latter offered a more straightforward way to evaluate the proposed systems in terms of algorithmic performance.
And, as often happens in scientific practice, the measure gradually became the objective~\cite{chen2018recsys}.

Attempts to bridge the gap between the evaluation paradigm and the original question have largely focused on introducing user studies that go beyond accuracy-based evaluation~\cite{knijnenburg2015evaluating, schedl2018evaluation}, or on addressing important adjacent issues such as biases, fairness, and algorithmic accountability~\cite{dinnissen2022fairness}.
Nonetheless, the underlying paradigm has remained largely unchanged for decades, and RS papers continue to be evaluated mainly through accuracy metrics on standard benchmark datasets.
Although critical perspectives have generally been welcomed within the community~\cite{cremonesi2021progress}, they have rarely succeeded in fundamentally transforming this evaluation framework.
Put simply, in the absence of a clear and practical alternative, the simplistic accuracy-driven paradigm has persisted.

A key reason why this paradigm is particularly limiting in the music domain lies in the intrinsic nature of music consumption itself. 
Unlike many other recommendation domains, music listening is deeply contextual, affective, and embedded in everyday activities. 
The same user may seek entirely different music depending on their momentary goal, such as focusing while working, regulating their mood, socializing with friends, or expressing their identity. 
For instance, a user might request "calm instrumental music for studying," "energetic tracks for a workout," or "songs similar to this but less melancholic." 
These needs cannot be fully captured by past consumption alone, as they reflect transient intentions, subjective descriptions, and contextual rather than stable long-term preferences. 
As a result, the next item predicted from consumption patterns may not correspond to the music that best satisfies the user’s current needs.

Music recommendation is further complicated by the nature of music items and the feedback they generate. 
Music is multimodal and the metadata used to describe music, such as genre, style, or mood, are often subjective, culturally dependent, and inconsistently defined.
For example, descriptors such as "relaxing," "dark," or "summer vibe" may refer to different musical characteristics depending on the listener and context.
At the same time, user feedback is mostly implicit, consisting of noisy behavioral signals such as play counts, skips, or listening duration, which provide only indirect and ambiguous evidence of preference. 
These properties make it difficult to rely on a single ground truth, to represent music and users precisely, and to evaluate recommendation quality using conventional metrics alone as they reflect challenging fundamental characteristics of music consumption instead of objective constraints.
This stands in contrast to domains such as movies or e-commerce, where preferences can be more easily specified using factual attributes such as actors, or objective product features.

The emergence of Large Language Models (LLMs) and their rapid adoption in RSs disrupts this long-standing evaluation practices, as the fundamental assumptions of the established paradigm no longer hold~\cite{deldjoo2025toward,deldjoo2024recommendation}.
LLMs are generative models: they do not compute similarity scores nor rank items in the conventional sense.
Therefore, they cannot be reliably evaluated using standard accuracy metrics.
More critically, such metrics were not designed to account for hallucinations, knowledge cutoffs, the influence of training data sources, or personalization.

These limitations are particularly consequential in the music domain, where recommendations must align with subjective contextualized NL requests or descriptors provided by users, such as mood, activity, or cultural context, resolve ambiguous artist or track names, and remain grounded in large, rapidly evolving, and long-tailed catalogs.
Moreover, the very notion of train/test separation becomes largely irrelevant for black-box models such as ChatGPT---especially given that these models have likely been exposed to most of the public datasets commonly used for RSs evaluation~\cite{10.1145/3726302.3730178}, making reproducibility and fair comparison of results difficult, if not impossible.

At the same time, LLMs have inspired a new wave of methodologies and can even be applied directly to the core MRS objective, that is, users can simply ask an LLM for music recommendations, much as they would ask a human acquaintance, or they could ask LLMs to assess recommendations in a role as judges.

\begin{figure}
    \centering
    \caption{Paper's overview as a generic diagram presenting music recommendation with LLMs}
    \includegraphics[width=0.95\linewidth]{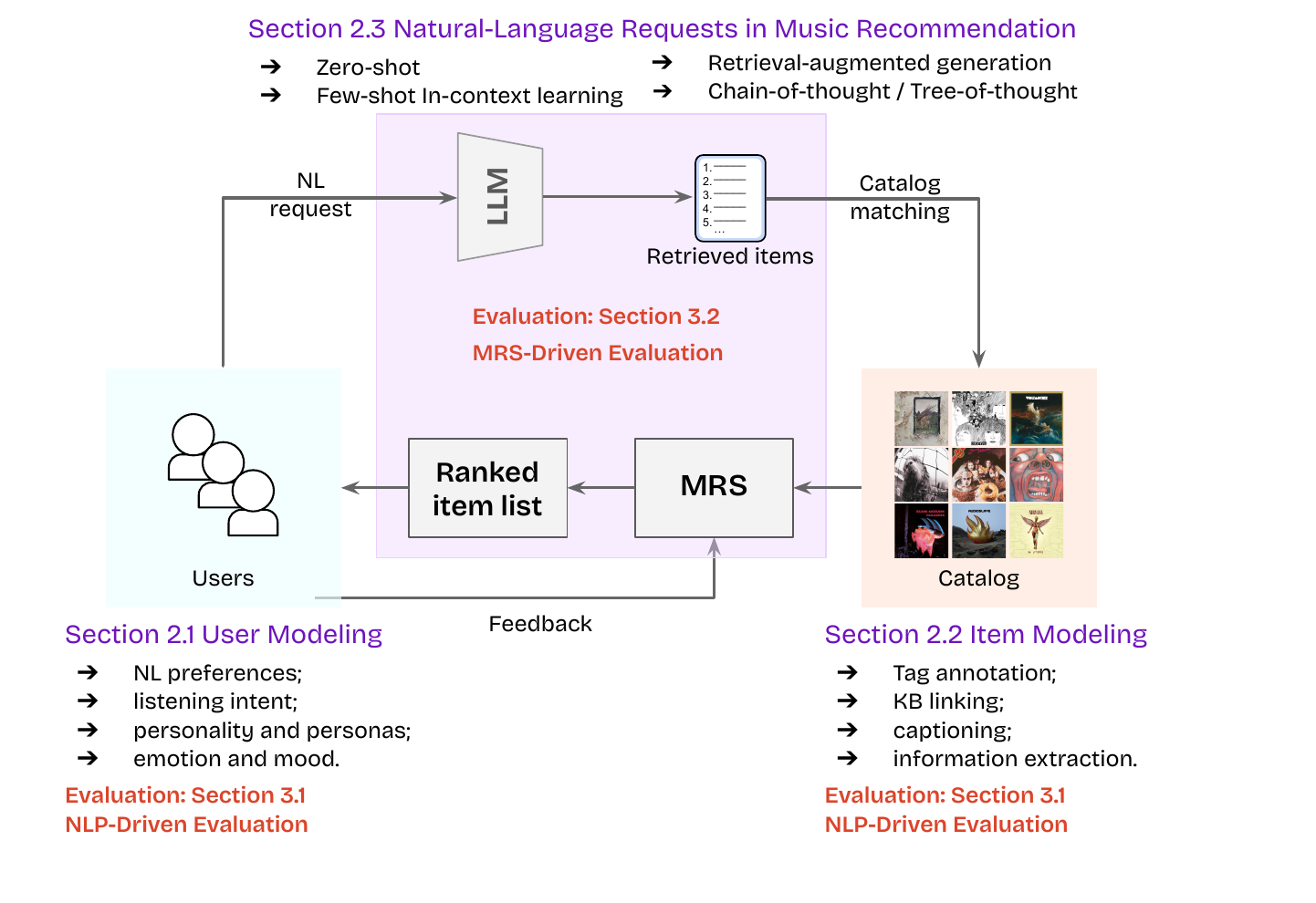}
    \label{fig:paper_overview}
\end{figure}

Despite the breadth of challenges that LLMs introduce to traditional MRSs evaluation, many works on LLM-based MRSs have continued to rely on standard accuracy-based metrics, often without fully considering what these numbers signify for models that generate text rather than predict or rank discrete items.
Regardless of where one stands in the debate over the disruptive nature of LLMs, their growing use in MRSs represents a valuable opportunity to revisit how we evaluate RSs.
It encourages the exploration of new evaluation perspectives that better align with the human experience of searching for and receiving music recommendations, while also prompting reflection on what constitutes a poor recommendation in generative settings, such as cases of hallucination or factual inconsistency. Effectively, it also requires taking inspiration from the field of Natural Language Processing (NLP) to evaluate the text produced by LLMs.

Against this background, we provide the article at hand, which constitutes a synergistic combination of survey and position paper, and is structured as follows:
In Section~\ref{sec:2}, we provide a comprehensive overview of how LLMs have transformed various aspects of user modeling, item modeling, and recommendation, particularly in the music domain. 
Then, in Section~\ref{sec:3}, we first discuss evaluation at the general level, by considering each scenario as a text generation task and exposing what is done in adjacent research domains. 
We also draw inspiration from existing work in music recommendation to outline current strategies for measuring both success and risk, while introducing a detailed set of evaluation dimensions and challenges that we argue should be considered in any study involving LLMs as recommenders.
Figure~\ref{fig:paper_overview} provides a detailed overview of the article's structure. 
The main contributions are:

\begin{itemize}
  \item A detailed overview of background, challenges, and LLM-induced changes in MRSs, with an emphasis on the specificity of music consumption (Section~\ref{sec:2}).
  \item An NLP-inspired evaluation taxonomy that goes beyond traditional IR metrics and covers a wider range of scenarios, including human vs.~automatic evaluation and reference-based vs.~reference-free settings (Section~\ref{sec:3.1}).
  \item A concrete MRS-oriented evaluation framework with operational success metrics, failure-side diagnostics, and working examples for LLM-based music recommendation (Section \ref{sec:mrs-eval})
\end{itemize}

\section{LLMs in Music Recommendation: Domain-specific Challenges and Opportunities}\label{sec:2}

In the following, we focus on three key aspects of MRSs where LLMs have been applied: user modeling (Section~\ref{sec:2.1}), item modeling (Section~\ref{sec:2.2}), and natural language-based music recommendation (Section~\ref{sec:nl_mrec}).
For each of these aspects, we structure the discussion around three main points: 
1) background or fundamentals of the task;
2) music-specific challenges and domain considerations, highlighting what makes the problem unique in the context of music;
3) impact of LLMs, analyzing how generative language models can or have already changed methods and capabilities in the respective task.

Past research on music item or user modeling for MRSs has mainly focused on two directions: 
(1) enriching items/users with explicit, often high-level information (e.g., tag annotation or extraction, knowledge base linking) and 
(2) representing items/users as features for use in content-based or hybrid recommender systems (e.g., representation learning). 
In what follows, we concentrate on the first aspect, which directly leverages the generative capabilities of LLMs to enrich the item or user information. 
The second aspect, involving text embedding and feature representation, is left for future work, as it encompasses broader approaches that extend beyond LLMs to include other text encoders and multimodal representation models.

\subsection{User Modeling}\label{sec:2.1}
A first use case of LLMs in the context of MRSs is to create and extend user profiles in an accurate and comprehensive way, tailored to the music domain and recommendation task at hand.
In the following, we first describe LLMs' potential for general music preference modeling, pertaining to traditional information such as consumption or interaction histories. 
Subsequently, we discuss their merits for advanced user modeling, involving personas, affective cues, listening intents, and cognitive effects.

\paragraph{Background.} Most automatic MRSs rely on modeling user preferences. These preferences are often represented as vectors or embeddings, learned from interaction data using neural architectures in collaborative filtering settings~\cite{kang2018self,he2017neural,hu2008collaborative}. A key limitation of such embeddings, derived from black-box models, is their opacity: they are not easily interpretable. This hinders both the generation of explanations for users and the investigation of the underlying decision processes.
For user representations to be effective, a comprehensive collection of interaction data is required, linking each user to a diverse set of items. This requirement introduces several drawbacks. For instance, cold-start users, those who have just joined the platform, lack a relevant representation and thus receive poor recommendations. Similarly, if a user wishes to steer their recommendations toward new interests, the system typically demands a substantial number of interactions with items from these new domains before the model can adapt.

User profiles can also span to the level of \textit{personas}, i.e., groups of users that share common characteristics or tastes. 
These can be defined, among others, at the level of demographics (age, gender, place of residence), involve cultural aspects (religion, beliefs, norms), or socio-economic factors (wealth, income, social class).
Together with models that connect one or more of these factors to music preferences, e.g.~\cite{zangerle_user_2020,liu_relation_2018}, such personas have been used to overcome cold-start scenarios. 
For instance, the MRS can only use the user's belonging to a demographic group, to make recommendations, in a traditional demographic filtering approach.
Alternatively, the definition of personas can also reflect more closely certain music-related patterns or dispositions. For instance, personas can be modeled as having different preferences for musical diversity, popularity, or mainstreaminess~\cite{bauer_global_2019,lesota_computational_2023}.

Several works have provided evidence for a strong connection between music preferences and \textit{psychological aspects}, including \textit{personality}~\cite{melchiorre_personality_2020,schedl_interrelation_2018}, \textit{listening functions} and \textit{intent}~\cite{lonsdale_why_2011,schafer_psychological_2013} as well as \textit{affect}~\cite{fuentes-sanchez_musical_2022,peintner_nuanced_2025,parada-cabaleiro_exploring_2023}.
The purposes or reasons why people listen to music have been studied intensively in music psychology literature~\cite{lonsdale_why_2011,schafer_psychological_2013}, resulting in the emergence of three major trajectories, i.e., mood regulation, self-awareness, and social relatedness.
\textit{Personality}, often represented via the Big Five Model (aka Five Factor Model), describes traits along the dimensions of openness, conscientiousness, extraversion, agreeability, and neuroticism~\cite{https://doi.org/10.1111/j.1467-6494.1992.tb00970.x}.
While personality can also be part of a persona description, it goes beyond common persona definitions, especially in the context of music because of its relationship with musical perception and preferences. For instance, correlations have been identified between personality traits and acoustical properties such as loudness or tempo~\cite{melchiorre_personality_2020}, genre preferences~\cite{doi:10.1177/0305735607070382}, and perceived emotion~\cite{schedl_interrelation_2018}.
Based on such insights, a few attempts to integrate personality information into the music recommendation process have been made~\cite{DBLP:conf/recsys/LuT18,DBLP:conf/jcdl/LiuH20,DBLP:conf/aaai/Shen0LMBWCCH20}.
Furthermore, modeling affective states of listeners is strongly related to modeling \textit{listening intents}, as mood regulation is one of the dominant reasons why humans listen to music~\cite{schafer_psychological_2013}.
Eliciting affective cues such as \textit{emotions} (short-lasting reactions in response to a stimulus), \textit{mood} (longer-lasting episodes independent of a concrete stimulus), or \textit{sentiment} (general positive or negative attitude towards a concrete item or item group) is thus important to tailor recommendations to the user's current affective state.

\paragraph{Music-specific challenges.}
Compared to domains such as books, movies, or e-commerce~\cite{ramos2024transparent,  gao2024end, zhou2024language}, music consumption presents a number of particular challenges for user modeling in the context of MRSs: 

\begin{enumerate}
    \item Music catalogs are extremely large, often spanning tens of millions of tracks. Moreover, music catalogs tend to be very rich in terms of culture, genre, year of production etc. As a result, users interact with hundreds or even thousands of unique tracks over time~\cite{schedl2018evaluation}. 

    \item On music streaming platforms, the most common form of feedback is implicit. Implicit feedback refers to user actions and behaviors that are not explicitly provided as signals of preference. In this setting, feedback is typically derived from listening time or from binary listening events defined by a time threshold, often set at 30 seconds. The challenge stems from implicit feedback being a noisier signal of interest than explicit ratings or reviews~\cite{sguerra2025uncertainty}. Moreover, in the absence of explicit textual feedback, user preference modeling must rely on item metadata, whose quality varies considerably~\cite{matrosova2024recommender}.

    \item User musical preferences are complex, often spanning both short- and long-term dimensions~\cite{sguerra2025biases}. Short-term preferences reflect recent interactions and can be quite noisy, while long-term preferences capture more stable tastes developed over years. However, long-term preferences may also become outdated and lack novelty. Moreover, musical preferences are highly contextual. For example, soundtracking, selecting music for a given activity, is a very common way of consuming music~\cite{fuentes2019soundtracking, marey2024modeling}. In different contexts, users will change the type of music they consume to match the mood or the energy pertaining to their current activity: someone will play energetic music when working out or calm music for studying. This can drastically change the user's preferences, something that a static user representation might not be able to pick up. 
    
    More broadly, soundtracking illustrates a functional use of music, where listening is guided by an explicit goal rather than by genre or artist affinity. In many everyday scenarios, music is used to support an intended outcome (e.g., focusing, regulating mood, or increasing motivation). These functional aspects of listening are consistent with evidence that music can influence human cognition, emotion, and behavior~\cite{rentfrow2012role}. Thus, music consumption is often highly situational, shaped not only by enduring taste but also by immediate purpose and context.

    \item Affective responses can be quite subjective in nature when it comes to music. So are their impact on music preferences.
    A key distinction in this regard is between perceived and induced emotions. Music can both convey an emotion and evoke an emotional response, and the gap between these two aspects is non-trivial. Perceived emotion refers to the emotion recognized as being expressed by the music (e.g., identifying a piece as ``sad''), whereas induced emotion corresponds to what the listener actually feels while listening, which can vary substantially across individuals~\cite{song2016perceived}. This inconsistency further complicates affect modeling, particularly in the absence of explicit user-provided affective feedback.
    
    An additional factor making the use of affective cues challenging is that music is a highly multimodal artifact~\cite{deldjoo_content-driven_2024} and different channels of delivery (e.g., audio, lyrics, album cover, video clip) may elicit different emotions.
    Moreover, emotions may dynamically evolve over the time a user listens to a music piece. How to acquire and integrate such emotional trajectories in the recommendation process is still a complex problem.
\end{enumerate}

\paragraph{Key changes introduced by LLMs.} 

With the introduction of LLMs, a new alternative to traditional opaque user embeddings has become viable: the use of natural language representations of user preferences~\cite{ramos2024transparent, zhou2024language, gao2024end, sabouri2025towards, sguerra2025biases}. In this setting, LLMs act as summarization devices, producing high-level representations of users’ preferences from collections of interacted items. 
The use of textual taste representations offers many advantages, particularly their interpretability. Because these profiles are understandable, users can see what the system ``knows'' about them, increasing transparency and building trust. Moreover, in cold-start situations, users can explicitly state their preferences, enabling the system to generate recommendations. Furthermore, if users are allowed to edit their profiles, they could take control of their recommendations by steering them toward new interests and away from content they no longer find relevant. 

\begin{figure}
    \centering
    \caption{Representation of the method to derive NL user preference profiles from consumption data.}\includegraphics[width=0.85\linewidth]{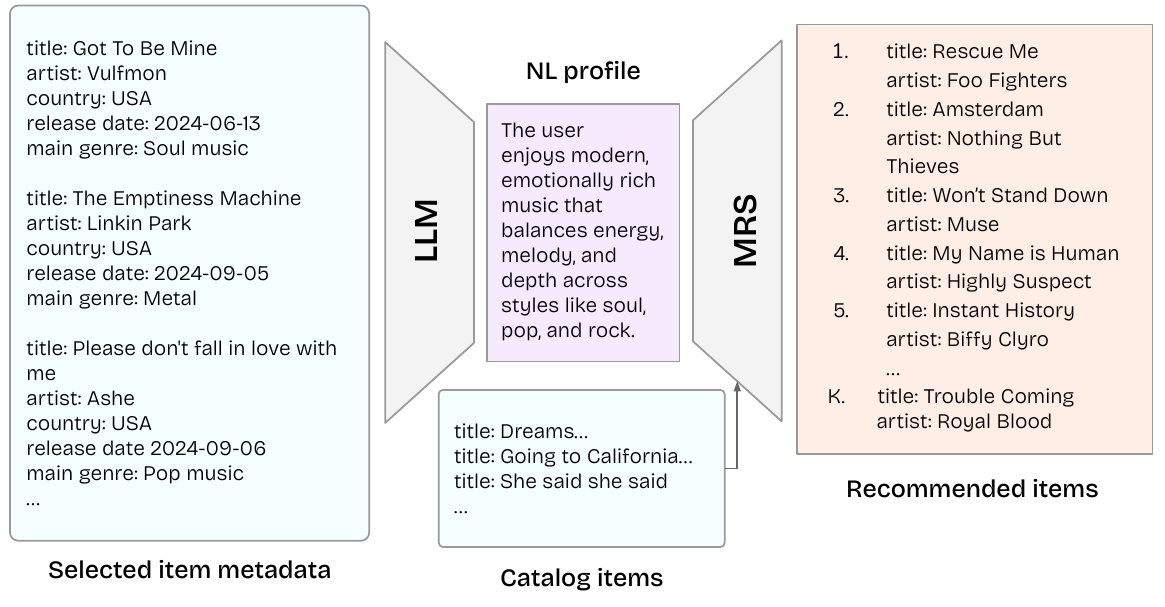}
    
    \label{fig:nl_profile}
\end{figure}

A common method for deriving natural language preferences with LLMs is to concatenate explicit feedback, such as item reviews, and metadata from consumed items into prompts and use the LLM to summarize this information into a profile~\cite{gao2024end,ramos2024transparent,zhou2024language,sguerra2025biases}. It is widely acknowledged that such profiles need to be concise in order to be scrutable. In this process, LLMs aggregate the most relevant information encoded in users’ consumption histories. In a second step, after profiles are generated or collected, recommendation can again proceed in multiple ways: either by projecting both user profiles and item metadata into a shared latent space~\cite{gao2024end,sguerra2025biases} or by employing a second LLM, employed here as recommender (described in Section~\ref{sec:nl_mrec}). A representation of this method is shown in Figure~\ref{fig:nl_profile}.

Given that LLMs are nowadays often perceived to exhibit human-like responses and show traces of "thinking" or reasoning, it is natural to investigate, and potentially leverage, the \textit{psychological traits} they may simulate, adapting concepts and mechanisms from human psychology such as personality, emotions, or cognitive effects and biases.
LLMs could enable new ways of approaching personality-related tasks in the domain of music recommendation.
For instance, they can be used to create personality profiles of their users (e.g., by predicting traits from the user's prompting history or by carrying out a conversation with the user guided towards inferring their personality using relevant personality assessment questionnaires)~\cite{zhang_rethinking_2025,sun_revealing_2024}. Enriched with correlation models between personality and music traits or preferences such as those proposed in~\cite{DBLP:conf/recsys/LuT18,melchiorre_personality_2020}, LLMs could then further tailor recommendations accordingly.
Also, LLMs appear to infer persona-related aspects such as musical diversity, popularity, or mainstreaminess~\cite{bauer_global_2019,lesota_computational_2023} from the user's listening history with some success.
Research on persona-based music recommendation can draw from recent large datasets of LLM-generated personas, e.g.~\cite{ge_scaling_2025}, that provide a synthetic dataset of 1 billion personas. 
While currently not tailored to the music domain, corresponding research can inform the creation of music-specific persona datasets, which in turn could be used to (1) support the classification of users into personas from their conversation with the LLM and (2) facilitate the simulation of music-related conversations with the system.

Regarding listening intents, LLMs provide a great tool to consolidate and extend these findings from music psychology.
For instance, integrating a data-driven approach with the results of the empirical questionnaire-based study from~\cite{schafer_psychological_2013}, Hausberger et al.~leverage LLMs to augment and cluster 129 listening functions into 32 broader intents and provide a simple interface that enables music search by intent~\cite{hausberger_exim_2025}.
Another pathway could be to infer the listening intent of the users from their conversations with an LLM, either through additional classifiers or directly through LLM-guided questions.
Ultimately, the listening intent could be used by traditional recommendation approaches, integrating it as contextual factor, or again directly by the LLM.

Surprisingly little research has been devoted to using LLMs for finding music that fits a specific \textit{emotion}.
Nevertheless, LLMs have shown some potential to identify emotions from conversations~\cite{fu-etal-2025-laerc,gao-etal-2024-cept}. Linking these to individual music preferences and using the resulting user model to tailor music recommendations, however, is still an open challenge with only very recent and limited research available~\cite{10.1145/3711896.3737212,wang_emotion-aware_2025}. 
This is a highly valuable direction given the importance of  emotion and mood in music perception and use. It is important to note that emotion and mood are different concepts from a psychological perspective, but often used interchangeably in the more technical RSs research works. Taking a psychological view, emotion refers to an affective response to a stimulus that is short-lasting but highly intense, while mood refers to a longer-lasting affective state of lower intensity that is not necessarily the result of a (single) stimulus.

\subsection{Item Modeling}\label{sec:2.2}
While collaborative filtering RSs primarily rely on user–item and item co-occurrence in streaming sessions, content-based or hybrid MRSs focus on a deeper understanding of an \textit{item’s features} during the recommendation process. 
These features align better with how humans perceive music, offering more support for explainable recommendations, enabling discovery, and item cold-start situations \cite{schedl2021music,deldjoo_content-driven_2024}.

\paragraph{Background.} Item features relevant to MRSs range from low-level attributes extracted from audio or lyrics to higher-level features such as user-generated content (e.g., playlist titles or music tags). 
\citet{deldjoo_content-driven_2024} propose an “onion model” to categorize musical content relevant for recommendation, placing \textit{audio} at the core, followed by \textit{embedded metadata} from music providers (e.g., title, artist, album, lyrics, label, recording year), and then outer layers of increasingly expressive, subjective, or culturally influenced features that reflect personal or community perceptions of music.  
In particular, \textit{expert-generated metadata} often focuses on musical aspects such as genre, style, or mood, but may also include details about music properties (e.g., instrumentation, vocal attributes, lyrical themes) or artist information like biographies. 
\textit{User-generated metadata} includes item descriptions such as tags, lyric interpretations, and music reviews, typically created on social media or forums (e.g., last.fm\footnote{\url{https://www.last.fm}}, reddit\footnote{\url{https://www.reddit.com/r/rock/}}, SongMeanings\footnote{\url{https://songmeanings.com}}) or within music streaming platforms when naming playlists or interacting with tracks after a search.

Embedded metadata and editorial information, such as genres, are typically available when music is ingested into streaming services. 
However, this data could be incomplete, ambiguous, or inaccurate. 
To tackle this, past works have explored matching musical items with knowledge bases or editorial encyclopedias, as well as aligning existing musical taxonomies or graphs to improve metadata verification, completeness, and music contextualization (e.g., adjusting music tagging taxonomies to a specific country) \cite{epure2019leveragingKB,epure2020modeling}.

Another approach to describing music, e.g., by its moods, styles, and instrumentation, is to rely directly on the audio signal. Many works in the Music Information Retrieval (MIR) community focus on tasks such as audio auto-tagging \cite{ibrahim2020should} or automatic audio classification \cite{rajan2021audio} (e.g., by genre, mood, or listening context); voice and instrument identification \cite{regnier2009singing}; lyrics transcription \cite{zhuo2023lyricwhiz}; and music captioning \cite{gabbolini2022data,gardner2024LLARK}.

When richer textual data related to music, such as reviews or music/artist descriptions, is available, established NLP techniques could be used for aspect extraction \cite{biancofiore2022aspect}, named entity recognition \cite{epure2023human}, and sentiment or negation identification \cite{rosa2015music}, as detailed in Table \ref{tab:extractive_tasks}.

\begin{table}[]
    \centering
    \caption{Extractive tasks for data augmentation when richer descriptions of music items are available.}
    \scriptsize
    \begin{tabular}{p{3cm}p{8cm}}
    \toprule
    \rowcolor{dzrblue!12} 
      \textbf{Extractive Task}   & \textbf{Description} \\ \hline
       \textit{Tag Extraction}  & Extract attributes or tags of music items from text (e.g., music genres, lyrical themes, instruments or voice characteristics) \\ 
       
    \rowcolor{dzrblue!12} 
       \textit{Named Entity Recognition} (NER) \textit{Coreference Resolution} & Detect names or mentions of music items (e.g., artists, albums, or tracks) from text. In creative domains like music, challenges stem from the frequent use of common words to denote named entities, and input text being often very noisy (e.g., user natural language queries or transcribed voice commands). \\

       \textit{Sentiment or Negation Identification} & Identify music items that are discussed with negative sentiment or explicitly disliked, which can be valuable for personalization and content filtering.
    \end{tabular}
    
    \label{tab:extractive_tasks}
\end{table}

\paragraph{Music-specific challenges.} Unlike domains such as books, news, or e-commerce recommendations, music content is largely non-textual apart from lyrics. 
However, lyrics are copyrighted, making their use difficult in a setting of reproducible research, as experimental datasets cannot be shared \cite{long2007insert,morea2005future}. 

In the context of music streaming services, textual information is often limited to short forms \cite{epure2019leveragingKB,epure2020modeling}, and common NLP techniques struggle with these short or noisy texts \cite{alsharou2021towards} (e.g., user tags or playlist titles, transcribed user queries).
When richer textual data is available, such as music reviews or social media posts, the text nature of track titles and artist names introduces additional challenges \cite{epure2023human}. 
Their vocabulary often overlaps with tagging terms. 
For example, "love" might refer to a track title, a mood, or a lyrical theme.
This ambiguity complicates key tasks such as named entity recognition and artist disambiguation (linking entities to the catalog or knowledge base).
Also, detecting negations as a way of expressing dislike (e.g., disliked music artists or genres) remains a significant challenge, even for LLMs \cite{garcia2023dataset,truong2023language}.

When relying on audio as input, music auto-tagging was until recently limited to fixed vocabularies, performing well for some tags but failing with contextualized or user-specific ones (e.g., music for "travel", "running", or "relaxing") \cite{ibrahim2020should}. 
Similarly, early audio captioning works, prior to transformer-based models, often produced unnatural or even grammatically flawed descriptions \cite{doh2021music}. 
Although newer models have improved linguistic fluency \cite{gardner2024LLARK,gabbolini2022data}, some studies show that audio is not always effectively leveraged as a modality \cite{weck2024muchomusic}. 

Finally, music, as a cultural product, is perceived differently across communities~\cite{celen2025expressions,epure2020modeling}. 
The same artist may receive different genre tags depending on the annotation practices of music providers or listeners, or the localization of Wikipedia editors. 
This variability is even greater for mood- or context-related tags, where personalization is essential but often absent (e.g., "party music" can mean very different things to different users \cite{melchiorre2025jam}). 
Consequently, aggregating and reconciling diverse knowledge sources for tag annotation, or music description generation in general, remains a significant challenge.

\paragraph{Key changes introduced by LLMs}

LLMs provide powerful capabilities for augmenting item information and have been leveraged in several complementary roles, each addressing different data-augmentation needs, as follows:

\textbf{LLM as Annotator or Knowledge Base.} Trained on vast amounts of online text, LLMs have become alternatives to knowledge bases or human annotators, even in zero-shot settings \cite{tekle2024music,bai2024KGQuiz,zhang2023llmaaa}. 
Thus, LLMs could generate user-friendly tags or keywords that describe a track, including its music genre or style.
LLMs could also be used to suggest editorial metadata such as release dates, country of origin, or artist details, when these are missing. 
Fine-tuned LLMs specialized in the music domain can go further by producing rich, descriptive content about items \cite{doh2024enriching}.
Compared to other knowledge bases, LLMs can offer a degree of personalization in generating music metadata or descriptions, whether through prompt manipulation (e.g., including characteristics of the target user or community), in solutions based on Retrieval-Augmented Generation (RAG) that interface with multiple tag taxonomies, or when used as personalized AI agents \cite{choi2025talkplaydata,lyu2024llm}.

\textbf{LLM as Information Extractor.} LLMs have proven highly effective at extracting specific information from unstructured text without even requiring a domain-adaptation or fine-tuning phase \cite{sainz2024gollie,hachmeier2025benchmark}. For example, given album reviews (e.g., Pitchfork\footnote{\url{https://pitchfork.com}}), social media posts (e.g., musicsuggestions subreddit\footnote{\url{https://www.reddit.com/r/musicsuggestions}}), or editorial content (e.g., rateyourmusic\footnote{\url{https://rateyourmusic.com}}), LLMs could identify and extract mentions of items or specific aspects associated with them. 
Moreover, LLMs could be directly integrated within conversational RSs and extract  examples of liked and disliked music as well as preferred and disliked musical attributes \cite{yun2025user} (e.g., "I like classical music but dislike repetitive vocals"). 
This process involves leveraging LLMs for Named Entity Recognition (NER), aligning an open vocabulary of short phrases with a fixed taxonomy, and high-level text reasoning to interpret expressions of dislike.
    
\textbf{LLM as Summarizer or Captioner.} LLMs are highly effective in generative tasks compared to any preceding models, particularly when the goal is to transform input data into new textual outputs with different structures or distributions from the source. 
In this role, LLMs are known to perform tasks such as: 
1) \textit{Summarization}---summarizing multiple sources of music item information into a concise, coherent summary or description; for instance, generating a short item caption from longer user descriptions focused on the interpretation of lyrics \cite{miyakawa2024emotion}. 
2) \textit{Item co-occurrence summarization}---analyzing patterns of music item co-occurrences and their metadata, then translating these insights into descriptive text that highlights relationships between items; an interesting application is generating smooth transitions or spoken segues between two tracks, similar to the commentary or explanations heard in radio shows \cite{Gabbolini2021Generating}.
3) \textit{Music captioning}--creating new content such as item captions or profiles, similar to the user profiles \cite{tekle2024music,doh2023lp}.

While all these roles and examples focus on textual data as input, \textit{Multi-Modal Language Models} (MMLMs) \cite{kong2024audio,deng2024musilingo}, which can incorporate other signals such as an item's audio, can also be leveraged to augment music items with additional information such as audio characteristics or album cover artwork. 

\subsection{Natural Language Requests in Music Recommendation}
\label{sec:nl_mrec}
We further focus on the setting where users express their music requests and preferences in natural language (NL). 
While communicating music preferences through language is arguably the most natural and socially grounded way of requesting music, resembling how people exchange recommendations among friends~\cite{yun2025user}, it remains a highly challenging scenario for modern Music Streaming Platforms (MSPs). Thanks to their strong textual understanding capabilities, LLMs have made this one of the most affected, and perhaps most promising application areas.

\paragraph{Background}
Given how natural it is for humans to articulate their needs in NL, MRSs with NL interfaces have emerged as a promising way for users to interact with the catalog~\cite{jannach2021survey}.
While most MSPs provide a search bar that allows users to make textual queries to access the catalog directly, their functionalities remain limited. The search bar represents a high-control interaction mechanism, enabling users to access specific content, for instance, music from a particular artist or album~\cite{sguerra2022navigational}. Users also rely on search when they only have a “seed of an idea” about what they are looking for~\cite{li2019search}, such as “music to work out” or “music for studying.” Despite the inherent limitations of traditional keyword-based search, users often adapt to its functioning and constraints over time.

The introduction of LLMs makes it possible to move beyond keyword matching or predefined frames in the case of traditional voice interaction~\cite{jannach2021survey}, and to support more complex and expressive user requests. 
Several studies have explored integrating them into recommendation interfaces that connect users’ NL requests to the catalog~\cite{delcluze2025text2playlist,oramas2024talking,melchiorre2025jam}. 
In this context, LLMs could draw on their extensive general “knowledge” of music to identify or describe relevant items.
Trained on massive and diverse sources of texts, predominantly from the web, they are able to capture and express musical attributes, such as tempo, rhythm, and timbre, through linguistic associations~\cite{carone2024soundsignature}.
This opens the possibility of bypassing traditional collaborative filtering pipelines and potentially mitigating issues such as popularity and exposure biases, which arise when models are trained solely on logged interaction data~\cite{matrosova2024recommender,ungruh2024putting,chen2023bias,huang2024going,pipergias2023collaborative}. However, LLMs also introduce a new set of challenges that must be carefully addressed.

\paragraph{Music-specific challenges}
Music makes for a particularly challenging recommendation scenario. It is a rich and multifaceted medium that spans cultures, countries, decades, and genres \cite{deldjoo_content-driven_2024}. 
As also discussed in Section \ref{sec:2.1}, music catalogs typically contain millions of tracks, each of which can be described through metadata such as release year, genre, or country of origin \cite{schedl2018evaluation}. In addition, the audio signal itself carries a wealth of information that can be extracted—such as mood, tempo, or instrumentation~\cite{peeters2004large, muller2015fundamentals}.
Music perception is both cultural~\cite{lee2025globalmood} and dynamic~\cite{sguerra2022discovery, sguerra2023ex2vec}, varying across individuals and communities. Moreover, music consumption is often highly contextual: as music influences human cognition, emotion, and behavior, it can be deliberately employed for specific goals such as improving concentration, maintaining motivation, or expressing social identity~\cite{rentfrow2012role}. Consequently, users often listen to music as a tool for emotional regulation, or selecting tracks that match a particular context or activity~\cite{marey2024modeling}.

These diverse dimensions make music an especially complex domain for NL requests. Users may ask for highly specific content (e.g., “Beatles songs from their early years”) or for abstract, affective experiences (e.g., “chill music with intense energy”). The challenge lies in bridging these linguistic expressions with the underlying catalog representations—while also accounting for users’ personal preferences, cultural background, and situational context.

Adding to this difficulty, new music tracks are continuously uploaded to the catalogs used by MSPs, creating persistent cold-start issues \cite{ferraro2019music}. 
When these new tracks are absent from the LLM’s training data, the model tends to favor older or more popular tracks, systematically overlooking newly released material. This is particularly problematic since many users rely on MRSs precisely to discover new music \cite{aluri2024playlist}.
Ultimately, NL-based music recommendation requires systems that can interpret language, not only semantically, but also affectively and contextually, aligning textual intent with both the structure of the catalog and the lived experience of music consumption.

\paragraph{Key changes introduced by LLMs.} LLMs enable more flexible and expressive user queries in music recommendation. Moreover, through prompt design, their behavior could be guided or constrained, for instance, to mitigate biases related to popularity, gender, or other social factors~\cite{deldjoo2024understanding}.
In the following, we detail the different ways LLMs have been employed as music recommenders. We focus on the scenario in which a \textbf{ranked list of musical tracks} is produced in response to a user’s NL request. This setting closely aligns with traditional top-K recommendation and playlist-generation tasks~\cite{delcluze2025text2playlist}.
We define the following task components:

\begin{figure}
    \centering
    \caption{Prompt example with the different task components highlighted.}\includegraphics[width=\linewidth]{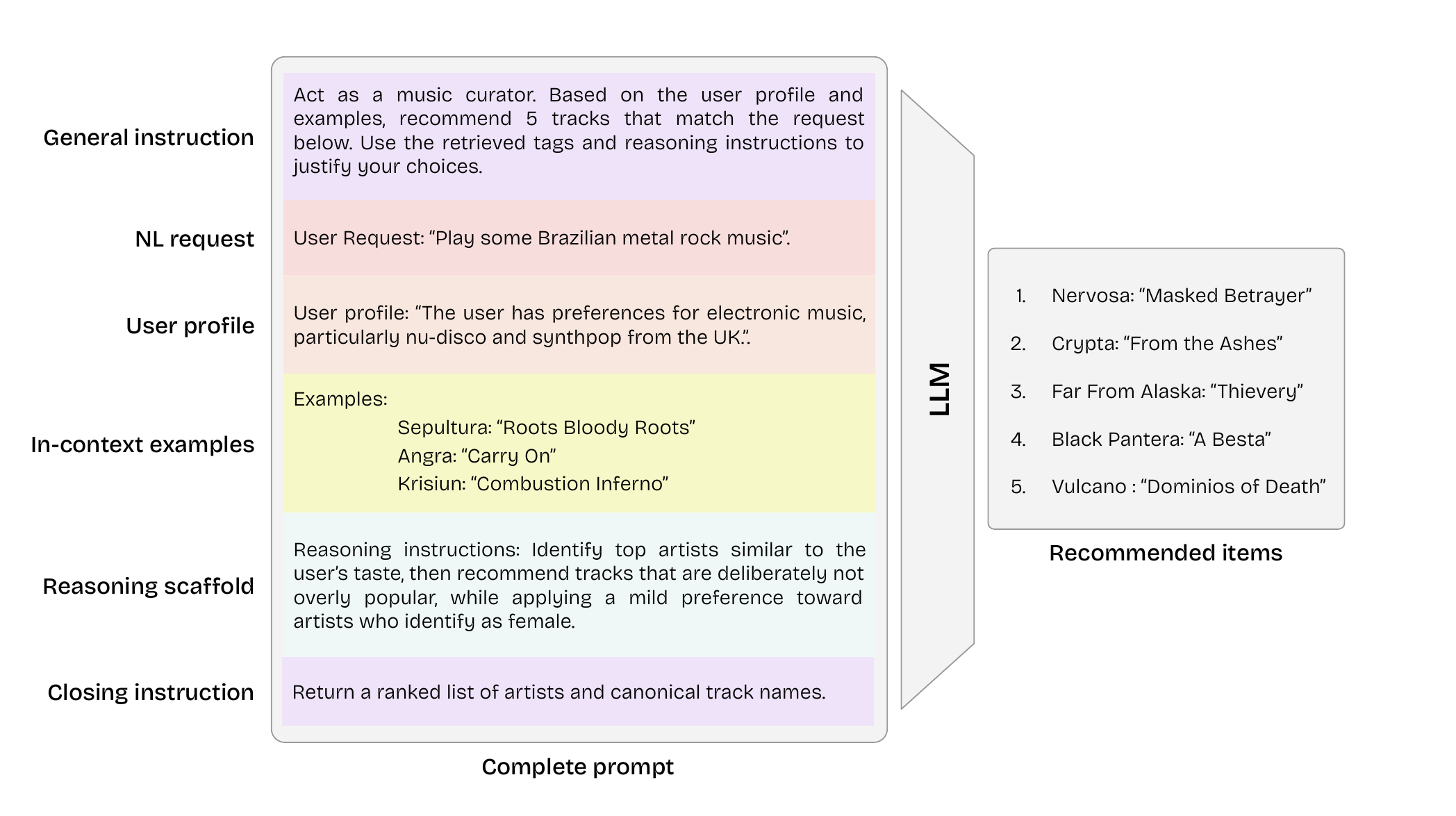}
    \label{fig:prompt}
\end{figure}

\begin{itemize}
    \item \textit{Music catalog:} track identifiers, canonical names, artist/album relations, and editorial or community metadata. Catalogs often contain multiple tracks sharing the same canonical names (e.g., homonymous artists, different versions of the same song across collections) and exhibit noisy or heterogeneous metadata, often supplied by labels, producers, or users;

    \item \textit{Resolver/knowledge base:} used for entity validation and disambiguation (e.g., MusicBrainz or the Music Ontology are often used for schema-level reasoning)~\cite{musicbrainz_homepage,raimond-2007-music};

    \item \textit{NL user request:} e.g., “play some obscure bolero guitar pieces,” “recommend me melancholic violin cues from anime soundtracks,” or “play Brazilian forró with accordion solos.”;
    
    \item \textit{NL user profile:} long- and short-term music preferences expressed in natural language~\cite{sguerra2025biases};
    
    \item \textit{In-context examples:} few-shot seeds / examples given as a list of tracks, artists, or other musical entities;
    
    \item \textit{Retrieved documents for RAG:} e.g., metadata snippets, tags, reviews, or new-release notices~\cite{lewis2020rag};
    
    \item \textit{Reasoning scaffold:} e.g., Chain-of-Thought (CoT) or Tree-of-Thought (ToT) prompting~\cite{wei2022chain,tot_yao2023}.
    
\end{itemize}
As outlined by the list of task components, we assume that generation is grounded in a \emph{resolver} or \emph{knowledge base} (KB) and may be further structured through prompting scaffolds. 
Specifically, as the MIR community has long emphasized domain-specific challenges such as \emph{entity disambiguation} (artist, release, recording)~\cite{ReganHristovaBeguerisseDiaz2023,musicbrainz_database}, \emph{heterogeneous metadata}~\cite{brooke2014descriptive}, and \emph{long-tail discovery}~\cite{celma2009music,ferraro2019music}, we explicitly separate the catalog, which is often noisy, from KBs, such as MusicBrainz\footnote{\url{https://www.musicbrainz.org}}, Discogs\footnote{\url{https://www.discogs.com}}, or AcousticBrainz\footnote{\url{https://www.acousticbrainz.org}}. The KB is then used to \emph{validate the existence of and disambiguate entities} (e.g., homonyms, release variants, remasters), and to map generated item names back to the catalog in order to filter out hallucinations.
Furthermore, a prompting operator composes these elements into a single prompt and a post-processor deduplicates, validates, and re-ranks the resulting list of tracks.
Figure~\ref{fig:prompt} illustrates a complete prompt, split into panes:
different recommendation tasks correspond to combinations of these elements. Next, we describe them in detail. 

\paragraph{Zero-Shot Recommendation (input: user NL request-only, possibly conditioned by user profile)}
In this setting, following the example in Figure~\ref{fig:prompt}, only the user request (e.g., “Play some Brazilian metal rock music”) is required, optionally accompanied by the NL user profile.

While classical RSs rely heavily on collected user feedback, zero-shot recommendation reframes the task as natural-language understanding, where the LLM maps open-vocabulary constraints to catalog items~\cite{brown2020language}. 
In the absence of a user profile, the model relies entirely on its linguistic priors.
LLMs internalize broad cultural knowledge and can parse open-vocabulary descriptions, including compositional constraints, say, in terms of affect (valence, arousal, and  dominance)~\cite{warriner2013norms}, key/mode, or content control (e.g., ``avoid harsh vocals'').

In profile-conditioned zero-shot recommendation, personalization is achieved by prepending a readable description of the user’s taste. 
The profile is interpretable, editable, and directly leverages behavior, unlike opaque latent vectors~\cite{wu2024survey}.
This ``profile-as-text'' approach enables controllable personalization without modifying model weights, allowing users or curators to make immediate corrections~\cite{wu2024survey}. 

Zero-shot recommendation is particularly valuable in \textit{exploratory} and \textit{cold-start} scenarios as it operates without requiring any historical user interactions, instead treating the recommendation task as a language understanding and generation problem~\cite{wu2024survey,brown2020language}. Recent research shows that zero-shot LLM-based recommenders are promising in handling long-tail items and under-described genres that lack sufficient behavioral data~\cite{hou2024large,he2023large}. However, the zero-shot approach carries inherent risks: without explicit grounding in retrieved documents or contextual information, these systems are susceptible to popularity bias and may generate plausible but incorrect recommendations (i.e., hallucinations)~\cite{ji2023hallucination}.

The fundamental challenge stems from the model's reliance on parametric knowledge, i.e., information encoded in its weights during pre-training. Recommendation quality thus depends critically on two factors: accurate entity resolution that correctly maps user queries to items in the KB, and robust post-processing to validate outputs. When the model's parametric knowledge becomes stale relative to catalog updates, the zero-shot setting becomes particularly vulnerable to hallucinations, especially for niche or newly released items that were not present during training~\cite{ji2023hallucination}.

Despite these limitations, zero-shot approaches enable innovative user experience patterns. For example, systems can implement editable ``taste cards'' that allow users to modify their preferences and immediately see updated rankings without requiring model retraining~\cite{mysore2023editable,gao2024end}. To harness this flexibility while mitigating risks, practitioners should implement stricter entity resolution mechanisms and post-validation procedures to prevent the system from generating confident but incorrect recommendations~\cite{ji2023hallucination,lewis2020rag}.
In summary, zero-shot recommendation is a powerful tool for generating diverse suggestions and supporting exploratory discovery, provided that evaluation frameworks prioritize grounding accuracy and maintain balanced exposure across the item catalog.

\paragraph{Few-Shot In-Context Learning (ICL) (input: user request and in-context examples, possibly conditioned by user profile)} 
The zero-shot approach can be extended by adding in-context examples, short listening histories or seed items, that capture user preferences.
These examples can be sourced in two primary ways: from the target user's own interaction history (\textit{personalized few-shot})~\cite{deldjoo2024understanding} or from similar users' profiles in a collaborative setting (\textit{collaborative few-shot})~\cite{zhang2025AdaptRec,yu-etal-2024-peprec,bao2024realtime}.
Collaborative approaches are particularly valuable for cold-start scenarios or enriching sparse user profiles. 

Recent work explicitly designs systems to select similar user histories as in-context examples, enabling LLMs to actively participate in choosing these high-quality demonstration examples~\cite{zhang2025AdaptRec}. 
For instance, given the example in Figure~\ref{fig:prompt}, the items ``Sepultura - Roots Bloody Roots'', ``Angra - Carry On'', and ``Krisiun - Combustion Inferno'' serve as demonstrations to nudge the generation toward similar curation.
Whether drawn from personal history or collaborative neighborhoods, the model leverages a few explicitly liked tracks or micro-histories to enable immediate personalization, well-suited for adapting to new or shifting tastes without re-training~\cite{bao2024realtime}. Few-shot ICL is particularly useful in session-based recommendation, where short-term intent can be efficiently reflected through a curated list of seed examples (see, for example, AdaptRec~\cite{zhang2025AdaptRec}).

However, selecting appropriate seed items requires careful consideration regardless of their source.
Consumption logs are multi-intent and dominated by implicit signals (skips, repeats). If seeds are sampled naively from the chronological tail, recent interactions may overshadow long-term preferences and inject noise~\cite{sguerra2025biases}. In collaborative settings, this challenge intensifies: selecting exemplar users whose preferences are genuinely similar requires careful similarity computation to avoid amplifying biases from noisy neighborhoods~\cite{zhang2025AdaptRec}. Cross-lingual aliasing and transliteration are also a common issue: if all seeds belong to a single language, the model may amplify that linguistic or regional bias unless the selection is balanced~\cite{epure2020modeling}. Moreover, research on session-based recommendation documents strong order sensitivity. ICL inherits this hazard unless the selected examples are stress-tested with shuffle, drop, or paraphrase perturbations~\cite{ludewig2018evaluation,quadrana2018sequence}.

Beyond personalization, ICL also opens new design levers for controllability. LLMs can interpret heterogeneous examples, such as track IDs, textual notes (e.g., ``no harsh vocals''), or micro-histories, and synthesize them with open-vocabulary constraints, effectively producing a local curation policy ``in context''~\cite{brown2020language}. Designers can encode fairness or exploration directly (e.g., balancing genders or regions in exemplar artists), a controllability lever that classical embedding-only collaborative filtering does not offer~\cite{wu2024survey}.  Because seed examples exist exclusively as textual tokens within the prompt context, practitioners can perform A/B testing on prompt templates, system roles~\cite{deldjoo2024understanding}, and instructions without requiring model retraining or parameter updates~\cite{brown2020language}. 

\paragraph{Retrieval-Augmented Generation (RAG) (input: user request and retrieved documents, possibly conditioned by user profile)}

RAG enriches the prompt with retrieved documents drawn from either the catalog or an external KB (e.g., item metadata records, artist profiles, release information, user reviews, social tags, or new-release announcements)~\cite{lewis2020rag}. 
Evidence retrieved from the catalog enables the model to cite specific release groups and track identifiers, which can subsequently be mapped back to the catalog for execution. 
The fundamental goal is to ground recommendations in factual evidence and mitigate hallucinations by combining parametric knowledge encoded in model weights with non-parametric access to up-to-date external information~\cite{gao2023retrieval}. 
This paradigm is increasingly adopted to ensure catalog freshness by leveraging current information from external sources, since new releases and re-releases appear continuously in music streaming platforms~\cite{doh2024enriching}.

Music catalogs present unique retrieval challenges that distinguish them from general-domain applications~\cite{schedl2014music}. 
The same song may exist in multiple versions---original recordings, remastered editions, live performances---and metadata is frequently inconsistent or multilingual across databases. 
To address such complexities, RAG systems might employ two complementary search strategies. 
First, exact matching through catalog identifiers and structural relationships efficiently retrieves known items with high precision. 
Second, semantic search, which compares the contextual meaning of text rather than surface-level keyword matching, discovers relevant music even when users describe requests informally or in other languages~\cite{wu2025clamp}. 
The system then validates that retrieved items exist in the catalog and resolves ambiguities, such as multiple artists sharing identical names or cover versions among different recordings.

Two practical considerations fundamentally shape RAG implementation in MRSs. 
First, computational and economic efficiency: including excessively from retrieved content in the prompt increases both token consumption and processing latency, requiring careful ranking and selection of only the most relevant information to maintain real-time response guarantees. 
Second, catalog freshness and dynamic updates: as retrieval indices reside external to the LLM's parametric weights, music streaming services can immediately add newly released tracks to search indices without model retraining or fine-tuning~\cite{lewis2020rag}. 
This decoupled architecture enables LLMs to effectively bridge informal user requests (e.g., ``upbeat workout music'') with formal catalog metadata (genre classifications, tempo measurements, energy ratings)~\cite{doh2024enriching}. 
When appropriately validated through entity linking and relevance filtering, this approach enables scalable, factually grounded recommendations across millions of tracks while maintaining responsiveness to catalog evolution~\cite{gao2023retrieval}.

\paragraph{Chain-of-Thought / Tree-of-Thought (CoT/ToT) (input: user request and reasoning scaffold, possibly conditioned by user profile)}

Reasoning scaffolds guide LLMs to decompose complex recommendation tasks into explicit intermediate steps. For example, a CoT prompt for creating a workout playlist might instruct: \textit{``Generate a 30-minute workout playlist by: (1) selecting high-energy tracks with BPM 140-160, (2) filtering for motivational lyrics, (3) ordering by increasing intensity, (4) ensuring smooth key transitions between adjacent tracks, and (5) justifying each transition.''} Rather than producing recommendations through a single opaque generation pass, the model articulates each decision stage, making the curation process transparent and auditable.

CoT structures reasoning as a linear sequence of decisions (e.g., ``filter by mood $\to$ constrain by era $\to$ match key/BPM $\to$ order for flow $\to$ justify transitions''), where each step's output feeds into subsequent stages~\cite{wei2022chain}. 
ToT extends this by exploring multiple reasoning branches in parallel and aggregating across paths through self-consistency mechanisms or voting, improving robustness when single-path reasoning proves brittle~\cite{tot_yao2023,wang2022selfconsistency}. Both approaches enable the model to reason explicitly over multiple constraints, candidate selections, ordering criteria, and transition justifications.

Applied to music recommendation, these scaffolds provide three key benefits. First, \textit{interpretability}: users and curators can inspect why each track was selected and verify reasoning against catalog metadata. Second, \textit{controllability}: policy constraints (e.g., regional quotas, content restrictions, fairness objectives) can be encoded directly into the reasoning instructions without model retraining. Third, \textit{multi-constraint handling}: the system can balance competing musical dimensions (mood, tempo, key, energy progression) through explicit deliberation rather than implicit optimization.

However, music-specific challenges require careful handling. Subjective tags, such as mood or energy, are culturally dependent and annotator-specific, and without explicit grounding in catalog data, models may construct plausible-sounding reasoning chains that rely on invented micro-genres, spurious correlations, or biases~\cite{jacovi2020faithful}. Even objective metadata such as key or instrumentation can vary across different recordings or remasters of the same composition. Then, effective playlist generation requires global sequence-level coherence (i.e., smooth transitions, narrative arc, energy flow) beyond individual track quality, necessitating evaluation at pairwise and playlist-level granularity~\cite{schedl2014music}.

Critically, reasoning scaffolds should remain decision-useful, improving selection quality and constraint adherence rather than producing verbose but uninformative justifications. When properly grounded in catalog data and validated against factual metadata, CoT/ToT enables scalable, interpretable music curation that supports complex multi-stage reasoning while maintaining transparency and controllability.

\subsection{Discussion}
\label{sec:discussion}
We provided a detailed analysis of the three main areas of MRSs where LLMs have been applied: item modeling, user modeling, and natural language-based music recommendation. 
For each area, we structured the discussion around how the task has traditionally been addressed, the challenges arising from the specific characteristics of the music domain, and the impact of LLMs in addressing these challenges and existing methods and capabilities.

We now take a step back and summarize the key music domain characteristics and their implied challenges for MRSs in Table~\ref{tab:music-challenges}. 
The table is organized into three parts corresponding to the three areas above. 
However, several challenges cut across multiple areas:
for example, the extreme size of music catalogs and the continuous influx of new content affect item modeling, user modeling, and natural language-based recommendation alike.
The presented domain-specific particularities highlight that music recommendation involves more than matching users to items: it requires handling subjective, contextual, multimodal, culturally grounded, and highly dynamic preferences, often expressed through ambiguous or underspecified natural language that needs to be mapped on an extremely large and noisy catalog.

The exposed particularities and their corresponding challenges specific to the music domain should not be viewed in isolation, as they are interconnected and often reinforce one another. 
For example, the affective and context-dependent nature of music consumption interacts with the multimodal and culturally grounded representations of music, making both user modeling and item modeling inherently situational and subjective. 
Similarly, limitations in metadata quality and textual availability compound the semantic gap between natural language requests and catalog entities, especially in large, long-tailed catalogs where many items lack reliable tags. 
Implicit feedback and evolving short- and long-term preferences make it difficult to disentangle stable tastes from contextual or exploratory behavior, further complicating both personalization and evaluation. 
These interrelations highlight that many core challenges in music recommendation emerge from the interaction between these areas, rather than from any single factor. 
This also explains why addressing one aspect (e.g., improving item representations) often has direct implications for other components.

\begin{table}[t]
\caption{
%\textcolor{RoyalBlue}
Particularities specific to the music domain and their corresponding challenges for MRSs. Challenges are highly interdependent, with user dynamics, item representations, and natural language interaction mutually influencing modeling and evaluation.}
\scriptsize
\begin{xltabular}
{
\linewidth}{@{} p{0.01\linewidth} p{0.30\linewidth} p{0.49\linewidth} p{0.14\linewidth} }
\toprule
\rowcolor{dzrblue!12} 
 & \textbf{Music-Domain Particularities} & 
\textbf{Implied Challenges for MRSs} & \textbf{Related Particularities} \\
\midrule
\multicolumn{4}{l}{\textit{User-centric modeling and dynamics}} \\ \midrule
\rowcolor{dzrblue!12} 
1.1 & Evolving short- and long-term preference dynamics &
New users provide little or no history, and their initial preferences may reflect transient exploration or highly specific, short-term interests rather than stable tastes. Preference modeling should account for noisy short-term exploration signals while balancing them with more stable, though potentially outdated, long-term preferences. & 1.2, 1.3, 3.1, 3.2
\\

1.2 & Strong context dependence (music listening during an activity, for mood regulation, as a self-identity expression) &
Frequent preference shifts demand context-aware and adaptive recommendation models; stable preference modeling is not sufficient; item modeling is tightly linked to the user's perception and context. & 1.3, 2.1, 2.6 \\

\rowcolor{dzrblue!12} 
1.3 & Highly affective and emotion-driven user demands &
Subjective and dynamic responses demand highly personalized user/item modeling and evaluation. & 1.2, 2.2, 3.4
\\ 

\midrule
\multicolumn{4}{l}{\textit{Item representation and content characteristics}} \\
\midrule
\rowcolor{dzrblue!12}
2.1 & Multimodal nature (audio, lyrics, metadata, cultural annotations, video) &
Effective recommendation requires the integration of heterogeneous music facets, with some becoming central depending on the user's listening context, goal, or affect. & 3.5, 2.3
 \\
 
2.2 & Cultural and linguistic variability, subjectivity and personalization of music descriptors &
Genre labels and music descriptors vary across cultures, regions, and communities, complicating personalization, contextualization, and fairness. Contextual tags such as ``party'' or ``relaxing'' are highly subjective and differ significantly across users. & 1.3, 2.6, 3.4
\\

\rowcolor{dzrblue!12} 
2.3 & Incomplete and conflicting metadata; limited expressiveness of audio-derived descriptors & Music annotations originate from multiple sources with differing standards, leading to inconsistencies and conflicts.
Editorial metadata is often static and missing for a large part of the music catalog, limiting its reliability for representing items and matching them to users. Auto-tagging systems have historically relied on fixed vocabularies, failing to capture user-specific concepts. & 3.1, 3.5, 2.2, 1.1
\\

2.4 & Limited access to music textual content &
Music is primarily non-textual, and lyrics are often unavailable due to copyright restrictions, limiting accessible and reproducible datasets including lyrics. & 2.1, 2.3, 3.5
\\

\rowcolor{dzrblue!12} 
2.5 & Short and noisy textual signals &
Available music-related text (e.g., playlist titles, tags, queries) is often brief, sparse, or noisy, limiting its descriptive power to represent music or users. & 2.2, 2.6, 1.1
\\

2.6 & Music entity ambiguity in natural language queries, rare explicit expression of negation &
Track and artist names often overlap with common words (e.g., ``Love'') or multiple music entities could share the same canonical name, complicating entity recognition and accurate linkage to catalog items. User dislikes are often expressed implicitly or through negation in natural language, which is difficult to detect reliably. & 2.1, 2.5, 3.5
\\

\midrule
\multicolumn{4}{l}{\textit{NL interaction and MRS evaluation}} \\
\midrule

\rowcolor{dzrblue!12} 
3.1 & Extremely large and long-tailed catalogs, continuous influx of new content &
Persistent item cold-start, strong popularity bias, and difficulty ensuring fair exposure for less-known or novel music. & 2.3, 2.4\\

3.2 & Implicit feedback (play counts, skips, listening duration); passive consumption &
Noisy preference signals that complicate user preference modeling and reduce the reliability of offline evaluation. & 1.1, 3.4\\

\rowcolor{dzrblue!12} 
3.3 & Sequential, session- or playlist-based consumption &
Quality goes beyond item-level relevance, depending on collection ordering, transitions, and coherence. & 1.2, 1.3
\\

3.4 & Lack of clear ground truth for "good" music recommendations &
Evaluation must extend beyond accuracy to discovery, diversity, and user experience; it is highly linked to the user's music listening goal, context, or affect. & 2.2, 3.2\\

\rowcolor{dzrblue!12} 
3.5 & Complex and multi-dimensional natural language interactions to varied music facets &
User requests range from precise factual constraints (e.g., specific artists or periods) to abstract, affective, or contextual descriptions (e.g., mood, energy, activity). MRSs need to interpret complex, open intent and map subjective language to multimodal and multi-cultural catalog entities. & 2.1, 2.2, 2.3, 1.2, 1.3 \\
\bottomrule
%\end{tabular}

\label{tab:music-challenges}
\end{xltabular}
\end{table}

\paragraph{Concluding remarks on LLMs for User Modeling in the Music Domain} LLMs introduce new ways to model users by enabling the construction of textual profiles from listening histories, onboarding selections, or conversational interactions, which is particularly valuable in the music domain where explicit feedback is scarce and cold-start is frequent. 
However, unlike domains such as movies or e-commerce, music preferences are highly contextual, affective, and dynamic, meaning that a single static profile is often insufficient to capture the diversity of a user's taste. 
As a result, LLM-based solutions for music must go beyond generic profile summarization and support context-dependent representations, for example through multiple profiles or conversationally inferred personas that reflect different listening situations, emotions, or intents. 

This illustrates that LLM-based approaches must be adapted to the specific structure of music consumption, where preferences shift frequently and depend on subjective and situational factors.
At the same time, general LLM-based personalization methods do not directly transfer to music because they rely heavily on textual signals that are often sparse, ambiguous, or noisy in this domain. 
Profile construction from prompts or metadata may amplify noise, overlook negative preferences, or inherit inaccuracies from incomplete or conflicting music descriptors, while the affective nature of music requires interpreting nuanced emotional language without oversimplification. 
Moreover, LLMs must balance short-term listening signals with long-term preferences to avoid overfitting to transient behaviors. 
These domain-specific constraints imply that LLM-based user modeling for music requires additional mechanisms for grounding, calibration, and context awareness, as well as adapted evaluation protocols, compared to more text-rich and stable recommendation domains. Looking beyond these prompt-based configurations, recent work is increasingly framing recommendation as an agentic process, involving autonomous information seeking and, in some cases, coordination among multiple agents, thereby extending LLM-based recommenders from single-step generation toward more deliberative and interactive decision-making \cite{lin2026autonomous,maragheh2026future,banerjee2026collab}.

\paragraph{Concluding remarks on LLMs for Item Modeling in the Music Domain}
LLMs have expanded item modeling in music by acting as annotators, information extractors, and summarizers, enabling the automatic generation of tags, metadata, and natural language descriptions from diverse sources such as reviews, social media, or editorial metadata. 
These capabilities are particularly valuable for music, where metadata is often incomplete, inconsistent, or missing, and where content is largely non-textual. 
However, unlike in most other domains, music items are multimodal, culturally situated, and frequently part of the long tail or newly released, limiting the effectiveness of general LLM approaches.
As a result, LLM-based item modeling in music often requires additional grounding mechanisms, such as retrieval augmentation or multimodal alignment, to ensure accurate and up-to-date representations, especially for less popular or new content.

More broadly, general LLM-based item modeling methods do not transfer directly to music because they may fail to capture essential audio, contextual, or affective properties, and may propagate cultural biases, generic labels, or hallucinated attributes. 
When aggregating multiple sources, LLMs may also introduce inconsistencies or over-represent dominant cultural perspectives, which can negatively affect downstream recommendation and personalization. 
Music description must remain adaptable to different users and contexts, requiring controllable and personalized item representations rather than static tags or summaries. 

Beyond subjective tags, even mid-level MIR descriptors such as key, tempo, or chord estimation remain unreliable, particularly for non-Western or underrepresented musical traditions, given the Western bias of most datasets \cite{fuentes2019tracking,gomez2024salsa,maia2022adapting, cano2021sesquialtera}. 
Thus, extracted features may be systematically inaccurate when applied to diverse repertoires. 
This issue is amplified in LLM-based pipelines, where generated descriptions often rely on upstream MIR outputs: 
if these outputs are flawed or culturally misaligned, the resulting captions may appear coherent yet misrepresent the music.

\paragraph{Concluding remarks on LLMs for Natural Language Music Recommendation} LLMs enable NL-based music recommendation by directly generating ranked item lists from user requests through prompting strategies such as zero-shot, few-shot in-context learning, retrieval-augmented generation (RAG), and reasoning scaffolds. 
These approaches allow flexible interaction, making them particularly promising for cold-start users, exploratory queries, and complex intents that combine factual constraints with subjective or affective descriptions. 
However, the effectiveness of these methods in music depends strongly on how prompts are constructed and grounded. 
Zero-shot and few-shot approaches rely heavily on the model’s prior knowledge or selected examples, which may over-represent popular, culturally dominant music or short-term  user interests. 
Moreover, in music, other challenges are the ambiguity of entity names, the subjectivity and cultural dependence of descriptors such as mood or energy, and the sparsity and inconsistency of textual metadata. 
As a result, entity resolution and careful example selection become essential design components, and naïve transfer of general LLM prompting strategies may amplify noise, bias, or recency effects present in implicit listening logs.

Domain-specific adaptations of LLM-based NL recommendation often incorporate retrieval grounding and reasoning scaffolds, which improve transparency, controllability, and factual alignment with catalog content. 
However, applying RAG in music can be challenging as relevant information is fragmented across heterogeneous and often inconsistent sources (e.g., metadata, tags, reviews), and only a limited portion can be included in the prompt.
Moreover, playlist generation introduces sequence-level constraints, where recommendation quality depends on transition coherence and overall flow. These domain-specific requirements mean that general LLM-based recommendation methods should not be directly applied without modification and careful consideration.

\section{LLMs in Music Recommendation: Evaluation}\label{sec:3}
In the previous section, we discussed the challenges associated with user modeling, item modeling, and NL-based recommendation in the music domain. 
We also summarized the key changes introduced by the adoption of LLMs in each of these areas. 
In what follows, we shift our focus to evaluating LLMs when used in any of the previously discussed scenarios.
%\textcolor{RoyalBlue}
\textcolor{black}
{We begin by providing a comprehensive overview of how LLMs are evaluated for generative tasks, as designed and discussed within the NLP community (Section \ref{sec:nlpeval}). We then extend the discussion to MRS-specific evaluation (Section \ref{sec:mrs-eval}), where we formalize universal success dimensions, discuss configuration-specific diagnostics, and illustrate the framework with working examples. Our goal is to provide an operational reporting framework and diagnostic toolbox for future empirical studies of LLM-based music recommendation.} 

\subsection{NLP-Driven Evaluation}
\label{sec:nlpeval}\label{sec:3.1}

When using LLMs for user modeling, item modeling, or music recommendation, the task generally reduces to learning $P_\theta(y \mid x)$, a conditional probability distribution over output texts $y$ (e.g., user preference profiles, playlist captions, music recommendations in natural language), given a textual or multimodal input $x$ (e.g., track metadata, text-based listening history, tokenized audio). 
For simplicity, the task can also be expressed as an input-to-text  mapping function as follows:
\[
f : X \rightarrow Y, \quad x \in X, \; y \in Y,
\]
where $X$ is the set of possible inputs and $Y$ is the set of possible output texts. 

\begin{figure}
    \centering
    \includegraphics[width=0.9\linewidth]{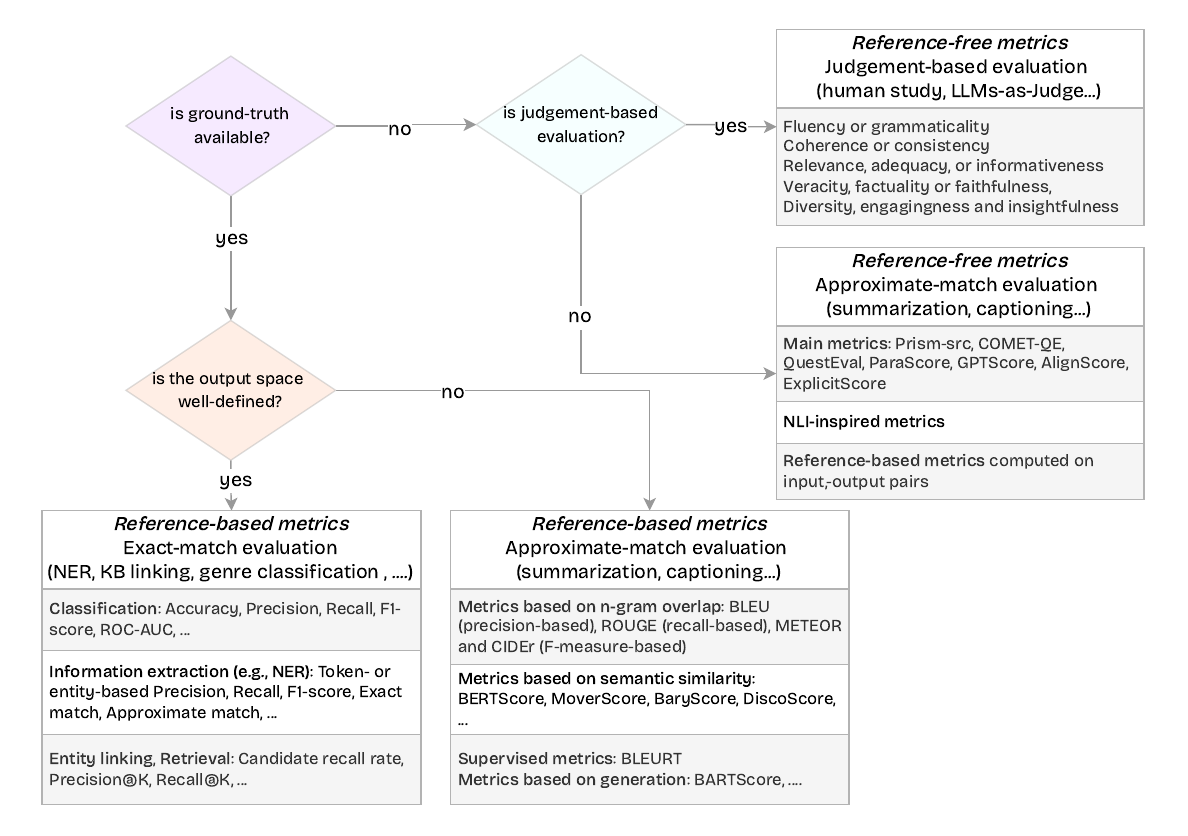}
    \caption{NLP-driven evaluation framework: decision process and metrics overview.}
    \label{fig:nlp-eval}
\end{figure}

Evaluation strategies can focus either on comparing the predicted output $y$ against a reference output $y^*$ or human judgments (\textbf{reference-based evaluation}), or on comparing the predicted output $y$ directly against the source input $x$ (\textbf{reference-free evaluation}). 
Reference-based evaluation, which \textit{relies on ground-truth data}, remains the most common approach. 
However, reference-free evaluation, often based on LLMs (e.g., LLMs-as-a-judge), has recently emerged as a promising, resource-efficient alternative, as it does not require extensive human annotation and is thus suitable for fast and large-scale evaluation experiments. 
Figure~\ref{fig:nlp-eval} summarizes the content that will be discussed in this section, presenting it as a decision process to guide the selection of the most suitable evaluation metrics based on the specific characteristics of each scenario.

In what follows, we set aside aspects related to content parsing. 
Modern LLMs can generate structured outputs directly (e.g., JSON files), which provides more suitable formatting of results for evaluation. 
Therefore, we do not consider evaluation metrics related to this parsing phase and focus only on the quality of the generated content itself.

\subsubsection{Reference-based Evaluation}
For the previously described roles of LLMs, the evaluation of success when references / ground-truth examples are available can be discussed as:
\begin{enumerate}
    \item \textbf{Exact Match Evaluation}, used when the output can be represented in a well-defined, discrete space with unambiguous ground truth (i.e., limited number of outcomes or outcomes expected to be limited to parts of input in case of information extraction).
    This includes tasks such as NER (i.e., matching predicted music entities to reference spans, either token-wise or entity-wise) \cite{epure2023human,hachmeier2025benchmark}, emotion annotation (e.g., music classification into multiple discrete classes \cite{yang2025exploring}), and knowledge base linking (e.g., checking whether a predicted item identifier corresponds exactly to the reference entity in the KB \cite{pond2025sesemmi}). 
    Formally, given a ground-truth output $y^*$ and a model-generated output $y = f(x)$, the \textbf{evaluation metric} can be defined as a function:
\[
M : Y^* \times Y \rightarrow \mathbb{R}, \quad  Y^{*} \subseteq Y \quad \text{and} \quad |Y^{*}| \ll |Y| 
\]
which measures the degree of similarity or correspondence between the reference and predicted outputs:
\[
M(y^*, y) \in \mathbb{R}.
\]
As $Y^*$ consists of a limited number of outcomes (e.g., a set of class labels, named entities, or knowledge base identifiers) in the exact match evaluation, $M$ corresponds to exact-match success metrics such as \textbf{Accuracy}, \textbf{Precision}, \textbf{Recall}, or \textbf{F1-score}. Ranking metrics such as \textbf{Precision@k} or \textbf{Candidate Recall Rate} could also be used, for instance, when assessing the linking of an entity to a knowledge base \cite{ruan2023entity}.

\item \textbf{Approximate Match Evaluation}, used when the output is free-form text or belongs to a large, open vocabulary
where multiple correct answers exist. Examples include music text summarization \cite{yuan2024chatmusician}, music captioning \cite{gabbolini2022data,gardner2024LLARK}, or user profile generation.
Formally, when $Y$ is the output space of open-form textual possibilities (i.e., any natural language text) with the ground-truth text $y^* \in Y$ for approximate match evaluation, the evaluation metric $M$ can be defined as a quality estimation function between the reference and predicted texts:
\[
M : Y \times Y \rightarrow \mathbb{R}, \quad
M(y^*, y) = S(\phi(y^*), \phi(y)),
\]
where $\phi(\cdot)$ denotes a transformation (e.g., embedding model, sparse feature extractor)
that maps text into a continuous representation, and $S(\cdot,\cdot)$ is either a similarity measure based on distance alignment, frequency-representation of n-gram overlap, or the probability of $y$ being generated from $y^*$ when estimated with another pre-trained generative model. 
In this setting, higher $M(y^*, y)$
values indicate closer semantic alignment between the reference and generated texts.
Typical choices of $M$ for this type of evaluation include: 

\begin{compactitem}

\item metrics based on n-gram overlap (\textbf{BLEU}~\cite{papineni2002bleu} — precision-based metric, \textbf{ROUGE}~\cite{lin2004rouge} — recall-based metric, \textbf{METEOR}~\cite{meteor2005automatic}, and
\textbf{CIDEr}~\cite{vedantam2015cider} — F-measure-based metrics), 

\item metrics based on semantic similarity (\textbf{BERTScore}~\cite{zhang2020BERTScore}, \textbf{MoverScore}~\cite{zhao2019moverscore}, \textbf{BaryScore}~\cite{colombo2021automatic}, \textbf{DiscoScore}~\cite{zhao2023discoscore}), 

\item  supervised metrics, i.e., regressive models on top of text encoder to predict human judgments (e.g., \textbf{BLEURT}~\cite{sellam2020bleurt}),

\item  metrics based on text generation (e.g., \textbf{BARTScore}~\cite{weizhe2021wide})

\end{compactitem}

\end{enumerate}
An overview of these evaluation metrics, exemplified when used for item modeling, is presented in Table \ref{tab:overview_metrics}.

Although widely adopted in the community, the approximate match evaluation metrics have been increasingly scrutinized. Studies have shown that n-gram–based metrics often correlate poorly with human judgments \cite{mathur2020tangled, peyrard2019studying}, while similarity-based metrics can be overly sensitive to adversarial perturbations or lexical variations \cite{sai2021perturbation}. 
To address these limitations, a new line of research \cite{chen2023menli, huang2025prompting} has proposed designing evaluation metrics inspired by the \textbf{Natural Language Inference} (NLI) paradigm, where a model-generated output $y$ is considered of high quality if it is semantically equivalent to a human reference $y^*$ under the notion of \textit{bi-implication} (i.e., logical equivalence). 
Furthermore, combining traditional metrics with NLI- or QA- (Question Answering) based approaches has been shown to yield more robust and reliable evaluation signals, improving correlation with human assessments  \cite{guo2024appls,chen2023menli}. 

\begin{table}[t!]
\caption{Overview of evaluation metrics, exemplified through item modeling tasks with LLMs.}
\small
\centering
\rowcolors{2}{dzrblue!12}{white}
\begin{tabular}{p{2.2cm}|p{4cm}|p{6.5cm}}
\rowcolor{dzrblue!12} 
\toprule
\textbf{Task Type} & \textbf{Example Output} & \textbf{Common Metrics} \\
\midrule
Classification & ``high valence'' / ``low valence'' & Accuracy, Precision, Recall, F1, ROC-AUC \\
Extraction (NER) & track=Love, artist=The Beatles & Precision, Recall, F1-score (token- or entity-level), Exact Match, Approximate Match \\
Entity Linking & Knowledge-based identifier: \texttt{track:12345} & Accuracy, Precision, Recall, F1, Candidate Recall Rate, Precision@k \\
Summarization & ``A warm, soulful album blending jazz and funk.'' & ROUGE, BLEU, METEOR, BERTScore, Human judgments (fluency, informativeness) \\
Captioning & ``A mellow jazz ballad with soft piano and vocals.'' & BLEU, ROUGE, CIDEr, BERTScore, Human evaluation (naturalness, relevance) \\ \bottomrule

\end{tabular}

\label{tab:overview_metrics}
\end{table}

\highlightbox{
Overall, the evaluation of natural language generation with the \textbf{reference-based framework} remains an active and evolving area of research. 
Current findings underscore the need to develop automatic success metrics that align more closely with human judgments, particularly across complex and multi-dimensional aspects of text quality. 
Moreover, there has been relatively little investigation into how general-domain evaluation metrics perform in specialized contexts, such as domain- or task-specific applications in RSs, where the nature of generated text and evaluation criteria may differ substantially.
In such settings, personalization, reflecting individual user preferences, tastes, and context, plays a crucial role in determining the perceived quality and relevance of generated content. This contrasts with most NLP text generation tasks, where personalization is typically less central to evaluation.}

\subsubsection{Reference-free Evaluation}
Because collecting human annotations for generative tasks is time-consuming and costly, a growing body of research has explored reference-free evaluation frameworks for tasks requiring an \textit{approximate match evaluation}, when the model’s output is evaluated with respect to the given input \cite{deutsch2022limitations,chen2023menli}.
In addition, these metrics are also considered valuable in online scenarios requiring real-time text quality estimation (e.g., captioning music playlists or answering questions about music in a conversation).
Formally, when $Y$ is the open-form output space (i.e., any natural language text, in the case of playlist captioning could be any music-related description), 
a reference-free evaluation metric $M$ can be defined as a function of the input $x$ (e.g., the metadata of the tracks in the playlist) 
and the model-generated output $y$ (e.g., the caption), which can be viewed as an explicit or implicit conditional 
generation model~\cite{thompson2020automatic,deutsch2022limitations}:
\[
M : X \times Y \rightarrow \mathbb{R}, \quad
M(x, y) = P_{\phi}(y \mid x),
\]
where $\phi$ denotes the parameters of the underlying evaluation model, 
and higher values of $M$ indicate better-quality outputs.
The underlying evaluation model $P_\phi$ is typically larger or more complex than the LLM used in recommendation, $P_\theta$, but it also has its own limitations and, therefore, cannot be used to solve the task directly---for instance, it may not be specifically optimized for that recommendation task or could be prohibitively expensive to run.
Reference-free metrics could employ evaluation models in a zero-shot setup (i.e., using pre-trained LLMs~\cite{fu2024gptscore}), 
while others~\cite{rei2021references} undergo an optimization process based on some training data to estimate $\phi$.

Examples of reference-free metrics, $M$, are \textbf{Prism-src}~\cite{thompson2020automatic}, \textbf{COMET-QE}~\cite{rei2021references}, \textbf{QuestEval}~\cite{scialom2021questeval}, \textbf{ParaScore}~\cite{shen2022evaluation}, \textbf{GPTScore}~\cite{fu2024gptscore}, \textbf{AlignScore}~\cite{zha2023alignscore}, \textbf{ExplicitScore}~\cite{chen2023exploring}, or \textbf{NLI-inspired} metrics \cite{chen2023menli}.
Reference-based metrics for text generation tasks (e.g., \textbf{BARTScore} or \textbf{BERTScore}) 
can also be adapted for reference-free evaluation by computing them between the input $x$ 
and the generated output $y$, provided $x$ is itself textual~\cite{shen2022evaluation}.

\highlightbox{
Although reference-free metrics were initially found to underperform compared to reference-based ones, 
recent work has shown substantial improvements in their ability to approximate human judgments of text quality~\cite{freitag2021results,shen2022evaluation}.
However, their reliability remains inconsistent across different natural language generation tasks, setups, and domains, 
with correlations to human evaluation varying considerably depending on  linguistic attributes, e.g., negation and sentiment polarity are poorly captured by embedding-based metrics in machine translation~\cite{freitag2021results}. Also, lexical divergence in paraphrasing---especially when text is written by humans---is measured inadequately \cite{shen2022evaluation}. 
This variability highlights an important research challenge: designing generalizable, 
reference-free evaluation methods that robustly capture human-perceived quality across diverse tasks, data distributions, and complex linguistic phenomena. 
Addressing this challenge requires a deeper understanding of how subjective dimensions of quality can be operationalized.
Furthermore, extending such evaluation metrics to handle \textbf{multimodal inputs}, 
as often encountered in music-related applications (e.g., captioning or cross-modal summarization), remains an even more challenging yet crucial direction for future investigation.
}

\subsubsection{Other Evaluation Approaches} 
As previously mentioned, \textbf{human evaluation} is widely regarded as the gold standard for assessing open-ended text generation tasks and automatic metrics seek to correlate with human judgments.
Human evaluation is also useful when automatic evaluation through metrics is not able to capture more subjective or complex aspects of quality.
Typical evaluation dimensions investigated in human studies of LLM-based MRSs include \cite{tsai2024leveraging,sayana2025beyond}:
\begin{itemize}
    \item \textbf{Fluency} or \textbf{grammaticality}, which measure how natural, readable, and error-free the output is;
    \item \textbf{Coherence} or \textbf{consistency}, which assesses if the text flows logically without contradictions;
    \item \textbf{Relevance}, \textbf{adequacy}, or \textbf{informativeness}, which reflects how well the generated text preserves important content from the source or fulfills the task;
    \item \textbf{Veracity}, \textbf{factuality}, or \textbf{faithfulness}, which analyzes if the output is accurate and grounded in the source material or external knowledge (this is also a measure of hallucination presence);
    \item \textbf{Diversity}, \textbf{engagingness}, and \textbf{insightfulness}, which captures originality and avoidance of repetition, as well as how interesting or likeable the output is (relevant in creative tasks such as music captioning).
\end{itemize}
Certain applications also employ task-specific criteria; for instance, \textbf{coverage and conciseness} in summarization \cite{khosravani2024enhancing}.

Quality judgments are often collected through Likert scales (e.g., 1–5 or 1–7) or pairwise preference rankings, in which annotators directly compare two outputs. 
Significant efforts have also been made to assess the correlation between human judgments on these complex dimensions and existing automatic metrics \cite{tsai2024leveraging}, with mixed results. 
For example, \textbf{coherence} has been shown to correlate relatively well with n-gram-based metrics (e.g., BLEU, METEOR), whereas \textbf{insightfulness} often shows no correlation with established metrics~\cite{tsai2024leveraging}. 

However, as we mentioned before, given that the evaluation involving humans still requires significant efforts for collecting results on these subjective dimensions, more recent works employ \textbf{LLM-as-a-judge} \cite{bavaresco2025llms,es2024ragas,chiang2023large,wang2024large}. 
LLMs appear to be reliable evaluators in a large range of tasks.
However, they also exhibit substantial inconsistency depending on the evaluated dimension (e.g., \textbf{engagingness} has low correlation with human judgment while \textbf{informativeness} correlates better), expertise level of human annotators (if compared with expert vs. non-expert human annotators), and if the assessed text is human- or LLM-generated \cite{bavaresco2025llms}. 

\highlightbox{Thus, LLMs should be rigorously validated against human judgments before being adopted as automatic evaluators. 
In the context of music and MRSs, this raises an open and important question: 
To what extent can LLM-based evaluators reliably and fairly assess 
music-related textual content, such as tags or generated descriptions, given the subjectivity and cultural variability inherent to musical preference and user perception? 
Addressing this question is required to ensure the reliability, bias control, and validity of LLM-driven evaluation in such personalized and domain-specific settings.}

\subsubsection{Discussion of Risks in LLM-generated Text Evaluation}

Multiple studies from the NLP domain \cite{rahnamoun2025multi,long2025llms,sun2022bertscore,chen2023menli,song2025good,miehling2025evaluating,rahnamoun2025multi} have outlined other pitfalls in evaluating LLM-generated text, besides those mentioned before. 
We identify risks stemming from the model itself and its intricacies, related to the characteristics of the generated output, and associated with the evaluation metrics.

\paragraph{Output-Related Risks}
In generative tasks, where the output can take many possible forms, recent research in NLP has identified several evaluation-related risks. First, the length of generated text has been shown to strongly influence evaluation outcomes for both n-gram-based and semantic similarity metrics~\cite{rahnamoun2025multi}. Because LLMs often generate substantially longer outputs than ground-truth references, automatic metrics may penalize otherwise relevant or high-quality generations due to length mismatches. Recent work shows that computing evaluation metrics on automatically summarized outputs, using simple extractive summarization methods, can significantly improve the alignment between automatic scores and human judgments.
Second, \citet{long2025llms} show that all LLMs have significant biases across various output formats (e.g., multiple-choice question formats, text wrapping styles, or list structures) spanning a wide range of benchmarks.
Several strategies to mitigate these biases have been proposed, including
\textit{(1) incorporating format templates or demonstrations in the prompt} and \textit{(2) repeating format instructions multiple times within the prompt.}

\paragraph{Metrics-related Risks} 
Risks pertaining to the used metric include social and representational biases, such as those observed with BERTScore \cite{sun2022bertscore}.
Reference-free metrics \cite{chen2023menli} also pose challenges: they can favor models similar to themselves, and thus, even penalize higher-quality human outputs, suggesting they are better suited as diagnostic tools instead of absolute measures of task performance.
Moreover, when LLMs are employed as evaluators, they have been shown to exhibit position bias \cite{wang2024large}.
{That is, the order in which candidate responses appear in the prompt can skew evaluation results, potentially making one model appear significantly superior to another.
To mitigate this bias, \citet{wang2024large} propose a calibration framework, which also shows closer alignment with human judgments.}

\paragraph{LLMs and Decoding Strategies} Given an input, the same LLM can produce highly variable outputs depending on the decoding strategy used (e.g., greedy decoding---selecting at each step the token with the highest predicted probability \cite{gu2017trainable}, or nucleus sampling---sampling from the smallest set of tokens whose cumulative probability exceeds a threshold \cite{Holtzman2020The}), a phenomenon referred to as non-determinism in generation \cite{song2025good}.
However, for practical and computational reasons, LLM evaluations are often conducted using a single input–output pair.
A recent study \cite{song2025good} evaluating multiple LLMs across several benchmarks has shown a significant gap between greedy decoding and sampling-based decoding:
greedy decoding often yields better results than the average performance obtained from multiple outputs generated through sampling strategies.
Also, employing a best-of-$N$ sampling strategy (selecting the best output among $N$ generations for the same input) can enable even smaller LLMs to match or outperform larger models.
{These findings carry important implications for the evaluation of LLMs.
First, the inherent non-determinism of generation introduces a source of variability that challenges the reliability and reproducibility of evaluation results: a single input–output pair cannot adequately represent model performance.
Second, decoding choices can significantly alter performance scores, making cross-model comparisons difficult unless decoding settings are standardized or explicitly reported.
Third, methods such as best-of-N sampling reveal that a model's potential quality can be much higher than what single-sample evaluations suggest, raising questions about what aspect of performance (average, best-case, or expected quality) should be measured.}

\subsection{MRS-Driven Evaluation}\label{sec:mrs-eval}
Music recommendation with LLMs inherits the evaluation challenges of both modern RSs and natural language generation while adding domain-specific complications such as extreme catalog scale and long-tailedness, frequent new releases, multilingual metadata, name ambiguity (artists/tracks that overlap with common words), and playlist-first consumption where ordering and transitions matter (Table~\ref{tab:music-challenges}, Section~\ref{sec:nl_mrec}). In this setting, classical ranking metrics alone are insufficient. 
Although non-existent or unresolved tracks should be penalized by the standard top-\(K\) metrics when they are evaluated directly against a fixed ground truth, aggregate relevance scores do not by themselves distinguish hallucinated items 
from irrelevant but valid items, nor do they capture failures that are removed by post-processing or hidden by averaging. An LLM-based system may thus, obtain acceptable relevance scores while still producing a catalog grounding failures, violating explicit constraints (``no live versions'', 
``only 1960s'', ``avoid explicit lyrics''), or systematically under-serving languages, regions, or music communities that are central to a listener's identity.

\subsubsection{Two Stakeholder Perspectives}
A central issue in evaluating LLM-based MRSs is that the notion of a ``good'' recommendation depends on the stakeholder under consideration. Evaluation should thus make explicit which utility the system is intended to optimize. Music streaming is a prototypical \emph{multi-stakeholder} setting: listeners, creators (artists), publishers (record labels), and platform providers often pursue different, and sometimes conflicting, objectives. These tensions are particularly visible in sequential listening and discovery-oriented scenarios \citep{dinnissen2022fairness,abdollahpouri2020multistakeholder,amigo2023unifying}.
The most commonly targeted stakeholders in research are listeners and creators. Thus, we detail their usual objectives next.

From a \textbf{listener} perspective, evaluation metrics are proxies for distinct facets of user value. First, \textbf{short-term satisfaction} asks if the system follows the user request \emph{in the current interaction}: it should return \emph{valid catalog entities} and respect explicit constraints (e.g., ``no live versions'', ``only 1960s'', ``avoid explicit lyrics''), since hallucinated or misattributed recommendations are failures even when the response is fluent \citep{ji2023hallucination,park2026echoes}. Second, \textbf{sustainable long-term value} (and discovery) ask if the system supports music discovery without extra focusing on overly popular content, reflecting the well-known exploration--exploitation tension between short-term satisfaction and long-term coverage, learning, novelty, and diversity \citep{kaminskas2016diversity,vargas2011rank}. Third, \textbf{personalization value}, \textbf{emotional well-being}, and \textbf{psychological effect} ask if recommendations feel genuinely tailored to the listener, add value beyond a one-size-fits-all or popularity-driven experience, and respond to affective needs and listening intents such as mood regulation, stress reduction, emotional expression, and social connectedness, thereby supporting overall well-being \citep{zhang2025impact,schafer2016goals}. Fourth, \textbf{agency}, \textbf{identity fit}, and \textbf{inclusion} ask if the system reflects user identity (e.g., language, scene, region, or cultural background) and provides sufficient controllability so that users can scrutinize and steer recommendations as tastes evolve \cite{radlinski2022natural,ramos2024transparent,hijikata2006content}. This point is particularly important for listeners with non-mainstream or minority preferences, who may otherwise be systematically underserved by popularity-biased systems \cite{kowald2020unfairness,khan2024experiences}. Finally, \textbf{trust} and \textbf{acceptance} are shaped not only by relevance but also by explanation quality: in music recommenders, explanations can improve trust, forgiveness, and user tolerance for unexpected suggestions, especially when recommendations are surprising but potentially valuable \cite{afchar2022explainability,tintarev2012evaluating}.

From a \textbf{creator} and \textbf{rightsholder} perspective, evaluation must also capture 
\textbf{attribution integrity} and \textbf{exposure balance}. Attribution integrity refers to 
if recommended works are correctly linked to the corresponding artists, tracks, albums, 
and rights holders. Exposure balance refers to if recommendation attention is not 
concentrated only on already popular artists, but also reaches long-tail, emerging, local, 
and culturally or linguistically diverse parts of the catalog. Recommendation is thus a 
form of attention allocation: as exposure is position-dependent, ranked recommendations 
can create downstream economic and cultural impacts for creators and provider groups. 
Fairness-aware evaluation should thus consider disparities across artist, label, provider, 
cultural, and linguistic segments \citep{singh2018fairness,dinnissen2022fairness}.

From a \textbf{platform-provider} perspective, evaluation should also report 
\textbf{operational} and \textbf{policy robustness}. Platforms operate under non-functional constraints, 
such as latency, cost stability, scalability, and catalog freshness, as well as governance and 
regulatory constraints, such as content moderation, safety, and compliance. These system-level 
concerns should be reported alongside user-facing effectiveness instead of being conflated 
with relevance or satisfaction metrics \citep{zangerle2022evaluating,zheng2023judging}. 

Because these objectives interact, we recommend \emph{multi-objective reporting} rather than 
collapsing evaluation into a single scalar score. For example, increasing long-tail exposure may 
improve discovery and creator-side exposure balance, but it may also reduce short-term hit-rate 
for some users. Similarly, stricter policy filtering may improve safety while reducing catalog 
coverage. Multi-objective reporting makes such trade-offs explicit and is better aligned with 
the inherently multi-stakeholder nature of music recommendation 
\citep{abdollahpouri2020multistakeholder,amigo2023unifying,deldjoo2024fairness}. More broadly, 
evaluation should ideally be informed by feedback from diverse affected parties instead of a single optimization target alone.

\input{final_2_1_mrs_driven_eval}

\section{Conclusion}
In summary, our work highlights how LLMs are already reshaping the development of MRSs. Their ability to interpret and generate text enables richer user and item representations and supports interaction modalities that extend beyond traditional recommendation pipelines. At the same time, we reiterate the importance of carefully adapting evaluation practices to these emerging capabilities. The generative, context-dependent nature of LLM-based systems challenges long-standing assumptions in MRS evaluation, reinforcing the need for methodologies that remain reliable, transparent, and aligned with real user behavior. As LLMs continue to evolve, ensuring that evaluation frameworks evolve alongside them will be essential for maintaining rigor and trustworthiness in RS research.

We posit that this is an opportune moment for the community to take a step back and reflect on evaluation practices that better capture the underlying goals of music recommendation. Beyond “solving’’ datasets, we argue for frameworks that address what constitutes a good music recommendation and how systems can support long-term user preferences, fairness, and healthier consumption patterns. Given the nature of LLMs, traditional performance metrics based on predicting consumed items become increasingly insufficient, both because of the opaque provenance of training data, which may overlap with popular recommendation datasets, and because such models can rely heavily on memorization to achieve high item-prediction scores. As we have described, the NLP community is comparatively more advanced in addressing these methodological concerns, and its practices can thus serve as a foundation for future MRSs and general RSs research to build upon.

We highly encourage further research on LLM-enhanced MRSs, to remain attentive to the risks these models introduce, and to design evaluation measures that reflect the goals and complexities of music recommendation. We expect our work to offer a grounded framework to navigate the opportunities, but also challenges of LLM usage, and the inherent subtleties of musical preferences and interactions that require careful consideration when building LLM-based MRSs.

\begin{acks}
This research was funded in whole or in part by the Austrian Science Fund (FWF): \href{https://doi.org/10.55776/COE12}{10.55776/COE12},  and  \href{https://doi.org/10.55776/DFH23}{10.55776/DFH23}, and \href{https://doi.org/10.55776/P36413}{10.55776/P36413}; and by the State of Upper Austria and the Federal Ministry of Education, Science, and Research through grant LIT-2024-13-SEE-111.
\end{acks}

\bibliographystyle{ACM-Reference-Format}
\bibliography{MRS_LLMs}


\end{document}

%% file: final_2_1_mrs_driven_eval.tex
\subsubsection{Universal Evaluation Dimensions and Metrics}
\label{subsec:success_overview}

We define success in LLM-based music recommendation as
\emph{faithfully satisfying a user request with real catalog items
while supporting discovery, personalization, controllability, and cultural reach}.
We introduce an MRS-oriented evaluation framework centered on multiple dimensions relevant across all LLM configurations, zero-shot, profile-conditioned, ICL-based, retrieval-augmented, or reasoning-based (Section~\ref{sec:nl_mrec}).
These dimensions answer six different questions:

\highlightbox{
\begin{description}
    \item[\textbf{G1. Query adherence and groundedness.}]
    Does the system return real catalog entities, and does it respect the user’s explicit constraints? This is the first and most minimal requirement for LLM-based recommendation, and it captures the \textbf{short-term satisfaction} aspect.

    \item[\textbf{G2. Discovery quality.}]
    When the user intent is exploratory, does the system promote discovery---for example, by increasing novelty and diversity and by allocating exposure beyond the head of the popularity distribution? This is a canonical ``beyond-accuracy'' objective, closely related to coverage and \textbf{long-term value}.
    
    \item[\textbf{G3. Personalization uplift.}]
    Does conditioning on the user profile or listening history improve recommendation
    utility compared with an otherwise identical non-personalized baseline, such as a
    generic, popularity-based, or request-only recommender? Here, ``uplift'' refers to
    the marginal gain attributable to personalization, for example a difference in
    nDCG@\(K\), QueryAdherence@\(K\), user-relative novelty, or human preference scores
    between profile-conditioned and profile-ablated recommendations.

    \item[\textbf{G4. Profile fidelity and controllability.}]
    Do recommendations match the user's stated preferences, and do they change \emph{predictably} when the user edits or refines their preference representation? This operationalises the \textbf{user-agency dimension}.

    \item[\textbf{G5. Cultural and linguistic coverage.}]
    Does the system represent the languages, regions, and music communities relevant
    to the request fairly and appropriately? Here, music communities refer to
    socially or geographically situated scenes or subcultures, such as local punk
    scenes, Detroit techno, UK grime, or regional metal scenes. This dimension
    captures if the group composition of the recommended list, for example its
    distribution over languages, regions, or music communities, matches the target
    composition implied by the user request or by a stated evaluation policy.

    \item[\textbf{G6. Classical relevance.}]
    How relevant are the recommended items under established top-$K$ effectiveness metrics? This dimension may become especially relevant when historical user interactions are converted into natural-language profiles to drive recommendations. It serves as the compatibility layer with standard recommender evaluation, but it should be interpreted jointly with G1--G5 in LLM-based MRSs.
\end{description}
}

The introduced dimensions are complementary. Running Example 1~(see Section~\ref{subsec:running_ex1}) shows that a system may be well grounded under G1 while still failing to satisfy the actual request. Running Example 2 (see Section~\ref{subsec:running_ex2}) shows that a system may achieve positive personalization gain under G3 while still failing discovery under G2. Accordingly, the role of the universal metrics is not to collapse evaluation into a single scalar score, but to expose distinct aspects of recommendation quality that matter in music recommendation. Table~\ref{tab:stakeholder-success-alignment} makes this explicit by showing how the listener- and creator/platform-side goals discussed above map onto the operational success dimensions G1--G6, while also clarifying that some system-level concerns, such as operational and policy robustness, should be reported alongside instead of being conflated with success metrics.

\begin{table*}[t]
\caption{Stakeholder-perspective interpretation of the universal success dimensions. The mapping is intentionally many-to-many: some goals are primarily captured by one dimension, whereas others require a joint reading of several dimensions. Operational and policy robustness remain important platform concerns, but they should be reported alongside rather than inside G1--G6.}
\label{tab:stakeholder-success-alignment}
\centering
\footnotesize
\setlength{\tabcolsep}{6pt}
\renewcommand{\arraystretch}{1.16}
\rowcolors{2}{dzrblue!12}{white}
\begin{tabularx}{\textwidth}{
>{\raggedright\arraybackslash}p{0.33\textwidth}
>{\centering\arraybackslash}p{0.15\textwidth}
X}
\toprule
\rowcolor{dzrblue!12}
\textbf{Stakeholder goal} &
\textbf{Primary dimension(s)} &
\textbf{Interpretation in the evaluation framework} \\
\midrule

\textbf{\textit{Listener}: request satisfaction and trust} &
\textbf{G1}, \textbf{G6} &
G1 checks if the system returns real catalog items, follows explicit constraints, and explains the list faithfully, while G6 adds the conventional top-$K$ relevance layer. Together they capture the minimal conditions for \emph{short-term satisfaction} and acceptance. \\

\textbf{\textit{Listener}: discovery and sustained value} &
\textbf{G2}, \textbf{G3} &
G2 measures long-tail exposure, lower popularity concentration, diversity, and catalog breadth. G3 complements this through user-relative novelty and the incremental benefit of profile conditioning beyond generic behavior. \\

\textbf{\textit{Listener}: personalization, agency, and affective fit} &
\textbf{G3}, \textbf{G4} &
G3 asks if profile conditioning adds measurable utility, whereas G4 asks if recommendations remain faithful to the user representation and respond predictably to edits thereof. For mood- or activity-sensitive requests, this is also where affective fit becomes visible. \\

\textbf{\textit{Listener}: cultural and linguistic fit} &
\textbf{G5} &
G5 evaluates if the distribution of the recommended list across languages, regions, and music communities aligns with the target implied by the user request or a stated evaluation policy. Here, music communities include scenes or subcultures such as local punk scenes, Detroit techno, UK grime, or regional metal scenes. The metric also assesses if failures are evenly distributed across such groups. \\

\textbf{\textit{Creator/platform}: attribution integrity and faithful explanation} &
\textbf{G1} &
Entity resolvability verifies that recommended items correspond to real catalog entities. Explanation--list consistency verifies that the rationale is faithful to the returned list and does not introduce unsupported attributes, artists, tracks, or reasons. \\

\textbf{\textit{Creator/platform}: exposure balance and ecosystem health} &
\textbf{G2}, \textbf{G5} &
G2 captures long-tail exposure, popularity concentration, and catalog breadth, while G5 captures representation balance across languages, regions, or scenes. Together, they address exposure allocation beyond pure relevance. \\

\textbf{\textit{Creator/platform}: operational and policy robustness} &
\textit{Reported alongside G1--G6} &
Latency, cost stability, safety, moderation, and regulatory compliance are important system-level concerns, but they are not themselves universal success dimensions in the present framework. \\

\bottomrule
\end{tabularx}
\end{table*}

Next, we operationalize the evaluation framework by making the dimensions explicit, defining the evaluation unit, and distinguishing strict feasibility constraints from graded satisfaction of fuzzy music descriptors. Several metrics are query-conditioned; their desired direction and interpretation should thus be assessed relative to the user intent and recommendation setting.

\paragraph{G1. Query adherence and groundedness.}
Let $u$ denote a user, $q$ a natural language request at a given conversational turn, $I$ the music catalog, and $R_K(q, u)=(i_1,\ldots,i_K)$ the top-$K$ items.
Unless otherwise stated, metrics can be computed per request and macro-averaged over an evaluation set $Q$.
Let $\mathrm{res}(i)\in\{0,1\}$ indicate if item $i$ can be resolved to a canonical catalog entity (e.g., a MusicBrainz identifier). We define:
\[
\mathrm{EntityGroundedness@K}(q)
=
\frac{1}{K}\sum_{j=1}^K \mathrm{res}(i_j).
\]
which measures the proportion of top-$K$ recommendations that can be successfully matched to entities in the target catalog, thereby capturing a fundamental validity requirement of free-text generative RSs~\cite{he2023large,deldjoo2025toward}. Higher values indicate that the system produces executable and catalog-grounded recommendations rather than fabricated or unusable outputs.
From the \textbf{listener side}, high values support \emph{short-term satisfaction} and \textit{trust}, since recommended items can actually be played or verified. From the \textbf{creator/platform side}, they help preserve \textit{attribution integrity} and \textit{reliable catalog coverage}. Conversely, low values point to hallucination-related failures (see below) that weaken usability, trust, and eventually long-term engagement~\cite{ji2023hallucination}.

We define the hallucination rate as the complement of entity groundedness:
\[
\mathrm{HallucinationRate@K}(q)
=
1-\mathrm{EntityGroundedness@K}(q).
\]

This metric measures the proportion of top-$K$ recommendations that cannot be resolved to entities in the target catalog, thereby capturing unsupported or non-existent recommendations~\cite{ji2023hallucination,deldjoo2025toward}. Lower values are desirable, since hallucinated items are one of the most visible failure modes of LLM-based recommendation.

We next separate strict query feasibility from graded query alignment. Let
$C(q)=\{c_1,\ldots,c_m\}$ denote the set of requirements to be evaluated extracted from query $q$. These requirements may include strict constraints, such as ``no live versions'' or ``released in the 1990s,'' as well as graded descriptors, such as ``not too mainstream'' or ``similar to this, but calmer.'' Let $C_{\mathrm{hard}}(q)\subseteq C(q)$ denote the subset of non-negotiable constraints. For each requirement $c_\ell$, let $s_\ell(i)\in[0,1]$ denote the degree to which item $i$ satisfies it, where $s_\ell(i)=1$ indicates full satisfaction. Following constraint-based recommendation and recent instruction-following evaluation, we first define the hard satisfaction predicate as~\cite{jannach2009constraintbased,abdin2024kitab}:
\[
\mathrm{Sat}_{\mathrm{hard}}(i,q)
=
\prod_{c_\ell \in C_{\mathrm{hard}}(q)}
\mathbf{1}\!\left[s_\ell(i)=1\right],
\]
with the convention that the empty product equals one when no strict constraint is specified. Strict query adherence is then defined as:
\[
\mathrm{QueryAdherence@K}(q)
=
\frac{1}{K}
\sum_{j=1}^{K}
\mathrm{res}(i_j)\,
\mathrm{Sat}_{\mathrm{hard}}(i_j,q).
\]

This metric measures the proportion of top-$K$ recommendations that are both executable catalog items and satisfy all non-negotiable query constraints. It is thus a strict feasibility measure: for a request such as ``Brazilian metal, no live versions,'' an item contributes positively only if it resolves to a valid catalog entity, is compatible with the requested style or origin, and is not a live recording. High values indicate strong request fidelity under explicit constraints, whereas low values indicate failures of grounding, constraint following, or both.

For subjective, fuzzy, or partially satisfiable descriptors, a strict binary predicate is often too severe. We thus define a weighted soft aggregation operator:
\[
A_{\alpha}\bigl(s_1(i),\ldots,s_m(i)\bigr)
=
\sum_{\ell=1}^{m}
\alpha_\ell s_\ell(i),
\qquad
\alpha_\ell\ge 0,
\qquad
\sum_{\ell=1}^{m}\alpha_\ell=1.
\]

The corresponding soft query alignment score is:
\[
\mathrm{QueryAlignment@K}(q)
=
\frac{1}{K}
\sum_{j=1}^{K}
\mathrm{res}(i_j)\,
A_{\alpha}\bigl(s_1(i_j),\ldots,s_m(i_j)\bigr).
\]

This metric measures the average degree to which the returned list matches the full set of query requirements, while allowing partial satisfaction. In contrast to \(\mathrm{QueryAdherence@K}\), it remains informative when the user request contains fuzzy musical descriptors, subjective affective terms, or trade-offs among multiple facets~\cite{jannach2009constraintbased,abdin2024kitab}. For instance, a track that matches the requested genre and decade but only partially matches the desired mood can still receive partial credit. The factor \(\mathrm{res}(i_j)\) prevents unresolved items from receiving alignment credit; a semantic-only variant can omit this factor, but should then be reported separately from catalog-grounded alignment.

To diagnose constraint-following independently of entity resolution, we also define the hard constraint hit rate:
\[
\mathrm{ConstraintHitRate@K}(q)
=
\frac{1}{K}
\sum_{j=1}^{K}
\mathrm{Sat}_{\mathrm{hard}}(i_j,q).
\]

This metric isolates if the recommended items satisfy the strict semantic requirements of the query, regardless of whether they can be resolved in the target catalog. Read together, \(\mathrm{QueryAdherence@K}\), \(\mathrm{QueryAlignment@K}\), and \(\mathrm{ConstraintHitRate@K}\) distinguish three cases: fully valid recommendations, near misses that partially match the intent, and failures caused specifically by catalog grounding.

Finally, let \(T_q\) denote the explanation text associated with query \(q\), and let
\(\mathcal{M}(T_q)\subseteq \mathcal{I}\) be the set of normalized catalog entities mentioned in that explanation. We define explanation groundedness as:
\[
\mathrm{ExplanationGroundedness}(q)
=
\frac{
\left|\mathcal{M}(T_q)\cap \{i_1,\ldots,i_K\}\right|
}{
\max\{1,|\mathcal{M}(T_q)|\}
}.
\]

This metric measures if the explanation refers to items that actually appear in the returned recommendation list, thus capturing a basic form of explanation faithfulness~\cite{jacovi2020faithfulness,tintarev2015explaining}. Higher values indicate that the rationale is grounded in the recommended list rather than introducing unsupported artists, tracks, or attributes. Low values signal a mismatch between the recommendation and its explanation, even when the list itself contains valid items.

\paragraph{G2. Discovery quality}

Let $\mathrm{pop}(i)\in[0,1]$ denote a normalized popularity score, and let
$\mathrm{LT}_\tau(i)
=
\mathbf{1}\!\left[\mathrm{pop}(i)\le\tau\right]$
indicate if item $i$ belongs to the long tail under threshold $\tau$, we define
\[
\mathrm{LongTailShare@K}(q)
=
\frac{1}{K}
\sum_{j=1}^K
\mathrm{LT}_\tau(i_j).
\]
which measures the proportion of top-$K$ recommendations drawn from the long tail of the catalog, following standard beyond-accuracy evaluations of popularity bias and exposure concentration~\cite{abdollahpouri2019managing,parktuzhilin2008longtail}. Within the discovery quality dimension, higher values indicate a stronger tendency to expose users to less popular and potentially under-discovered content.
From the \textbf{listener side}, high values suggest greater opportunity for exploration beyond overexposed content, while from the \textbf{creator/platform side}, they indicate broader long-tail exposure and a fairer distribution of attention across the catalog.
Importantly, the interpretation of this metric remains query-dependent: high long-tail exposure is particularly desirable in exploratory or discovery-oriented scenarios, but may be less appropriate for requests explicitly targeting canonical, trending, or mainstream content.

We then define the average recommendation popularity as:
\[
\mathrm{ARP@K}(q)
=
\frac{1}{K}
\sum_{j=1}^K
\mathrm{pop}(i_j),
\]
This metric measures the average popularity level of the recommended items and therefore provides a direct indicator of popularity concentration and head-catalog exposure~\cite{abdollahpouri2019managing}. Within the discovery quality dimension, lower values generally indicate a greater tendency to recommend less dominant and potentially less overexposed content.
For example, a recommendation list composed primarily of globally popular or canonical tracks will yield a high \(\mathrm{ARP@K}\).
From the \textbf{listener side}, lower values tend to support exploration and reduce repetitive mainstream exposure. From the \textbf{creator/platform side}, they indicate a less concentrated distribution of exposure across the catalog. Conversely, high ARP, especially when combined with low long-tail share, usually signals weaker discovery quality, reduced catalog breadth, and more concentrated creator-side exposure.

Let $d(i_a,i_b)\in[0,1]$ denote a pairwise dissimilarity measure between items $i_a$ and $i_b$. We define intra-list diversity as the average pairwise dissimilarity among the recommended items, according to some dissimilarity measure computed on feature representations of $i_a$ and $i_b$:
\[
\mathrm{ILD@K}(q)
=
\frac{2}{K(K-1)}
\sum_{1\le a<b\le K}
d(i_a,i_b).
\]
This metric is the standard list-level diversity measure used in RSs evaluation~\cite{ziegler2005diversification,vargas2011rank}. Higher values indicate that the recommendation list covers a broader range of items.
For example, a recommendation list spanning different artists, moods, genres, or audio characteristics will typically exhibit higher ILD than a list composed of closely related tracks.
From the \textbf{listener side}, high ILD reduces redundancy and helps keep exploratory sessions engaging. From the \textbf{creator/platform side}, it broadens exposure across different parts of the catalog. At the same time, ILD should still be interpreted jointly with relevance, so that diversity is not achieved simply by drifting away from the user’s request.

For system-level metrics, let
\[
\mathcal{Q}=\{e_n\}_{n=1}^{N},
\qquad
e_n=(q_n,u_n,\mathcal{Y}_n^\star,C(q_n),\pi(q_n)),
\]
denote an evaluation set of recommendation cases. Here, $q_n$ is the natural-language request, $u_n$ is the associated user when personalization is evaluated, $\mathcal{Y}_n^\star$ denotes optional ground-truth relevant items or relevance labels, $C(q_n)$ is the set of constraints extracted from the request, and $\pi(q_n)$ is an optional target group distribution used for cultural or linguistic coverage metrics. For reference-free metrics, $\mathcal{Y}_n^\star$ may be empty.

For an evaluation set $\mathcal{Q}$, we define system-level catalog coverage as
\[
\mathrm{CatalogCoverage}@K(\mathcal{Q})
=
\frac{
\left|
\bigcup_{e_n\in\mathcal{Q}}
R_K(q_n,u_n)
\right|
}{
|I|
}.
\]
which measures the fraction of the available catalog that is surfaced by the recommender across a set of requests, following standard beyond-accuracy evaluation in RSs~\cite{herlocker2004evaluating,vargas2011rank}. Higher values indicate broader catalog utilization and reduce the risk that the system repeatedly concentrates recommendations on a narrow subset of highly exposed items.
For example, if a recommender repeatedly cycles through the same small set of artists, catalog coverage will remain low even when individual recommendation lists appear relevant.
From the \textbf{listener side}, higher coverage supports \emph{long-term satisfaction} by reducing stagnation and enabling broader exploration. From the \textbf{creator/platform side}, it indicates healthier exposure across the inventory and less concentration of attention. Conversely, low coverage points to weaker discovery and more limited catalog use over time.

\paragraph{G3. Incremental value of personalization}

Let $U(\cdot)$ denote a utility functional (e.g., $\mathrm{nDCG@K}$, $\mathrm{QueryAdherence@K}$, or a human preference score).  
$R(q,u)$ denotes the recommendation list generated with user profile conditioning, while $R(q,\emptyset)$ denotes the corresponding non-personalized recommendation list generated without user-profile information.
For user $u$, we define the profile uplift as:
\[
\Delta_{\mathrm{profile}}(u)
=
\frac{1}{|Q_u|}
\sum_{q\in Q_u}
\left(
U(R(q,u))
-
U(R(q,\emptyset))
\right),
\]
where $Q_u$ is the evaluation set associated with user $u$.
This metric measures the incremental value contributed by profile conditioning rather than recommendation quality in isolation, and thus serves as a central metric for the personalization value dimension~\cite{herlocker2004evaluating}.  
From the \textbf{listener side}, a positive uplift means that the system contributes genuine personalization value. From the \textbf{creator/platform side}, it suggests better audience matching and less wasted exposure. Conversely, low or negative values indicate weaker personalization value, lower perceived care, and potentially reduced trust and long-term engagement.
We further define the inter-user overlap as:
\[
\mathrm{InterUserOverlap@K}
=
\mathbb{E}_{u\neq u'}
\left[
\frac{
|R_K(q,u)\cap R_K(q,u')|
}{
|R_K(q,u)\cup R_K(q,u')|
}
\right],
\]
where the recommendation lists are interpreted as sets when computing the Jaccard overlap.
This metric measures the average overlap between recommendation lists across different users, thereby diagnosing personalization collapse and insufficient user differentiation~\cite{jaccard1901distribution,kunaver2017diversity}. Lower values generally indicate stronger personalization, since they suggest that the system adapts recommendations to individual user characteristics instead of repeatedly returning the same items to everyone.
For example, if two users receive identical top-$10$ recommendation sets, then their pairwise overlap is equal to \(1\).
From the \textbf{listener side}, lower overlap suggests that recommendations are less generic and more identity-specific. From the \textbf{creator/platform side}, it supports more precise audience allocation and broader differentiation of exposure. Conversely, persistently high overlap indicates weaker personalization value and poorer identity fit, even when the returned lists appear superficially plausible.

Next, let $H_u\subseteq I$ denote the prior exposure history of user $u$ (preferably exposure rather than consumption alone in implicit-feedback settings).
We define user-relative novelty as:
\[
\mathrm{UserNovelty@K}(q,u)
=
\frac{1}{K}
\sum_{j=1}^K
\mathbf{1}[i_j\notin H_u].
\]
which measures the proportion of recommended items that are new relative to the user's prior exposure history, corresponding to the standard user-relative notion of novelty in RS evaluation~\cite{vargas2011rank,herlocker2004evaluating}. Higher values indicate that the recommender surfaces content that extends beyond the user's previously encountered items.
From the \textbf{listener side}, high novelty supports discovery and sustained engagement.
 From the \textbf{creator/platform side}, it broadens exposure to content that is new to the user instead of repeatedly reallocating attention to already familiar items. 
 Conversely, low novelty gain suggests that even if profile conditioning improves other metrics, the recommender may still fail to expand the user’s effective music space and thus contribute only weakly to \emph{long-term engagement}.

\paragraph{G4. Profile fidelity and controllability.}
For profile-based metrics, let $H_u$ denote the interaction history of user $u$, and let
\[
R_K(q,u)=(i_1,\ldots,i_K)
\]
denote the top-$K$ recommendation list returned for request $q$ and user $u$.
Let $\Phi(H_u)$ and $\Phi(R_K(q,u))$ denote smoothed facet distributions induced by the user's interaction history and by the recommendation list, respectively, over a fixed facet space $\mathcal{F}$, such as genre, decade, language, or mood. Thus, $\Phi(\cdot)$ maps a history or recommendation list to a normalized distribution over facets. Let $\mathrm{JSD}_2(\cdot,\cdot)\in[0,1]$ denote the base-2 Jensen--Shannon divergence. We define profile alignment as
\[
\mathrm{ProfileAlignment}@K(q,u)
=
1-
\mathrm{JSD}_2
\left(
\Phi(H_u),
\Phi(R_K(q,u))
\right).
\]

This metric operationalizes profile fidelity as calibration between the user's facet distribution of historic interactions and the facet distribution induced by their recommendation list, following prior work on calibrated recommendation~\cite{steck2018calibrated,lin1991jsd}. Higher values indicate that the recommended items more closely reflect the user's stated or inferred preference profile across the selected facet space.
For example, if the recommendation list closely matches the user's genre, language, and decade distributions, then \(\mathrm{ProfileAlignment@K}\) approaches \(1\).
From the \textbf{listener side}, high values support personalization, identity fit, and trust in the recommender. 
From the \textbf{creator/platform side}, they improve audience matching and reduce irrelevant exposure. Conversely, low values suggest weaker personalization value and poorer profile fidelity, and in requests involving mood or activity, they may also indicate reduced affective appropriateness.

Let $e$ denote an explicit user edit (e.g., ``same mood, but instrumental''), and let $q\oplus e$ denote the modified recommendation request obtained after applying the edit. We define edit success as the proportion of returned items that satisfy the updated constraint set:
\[
\mathrm{EditSuccess@K}(e)
=
\mathrm{ConstraintHitRate@K}(q\oplus e).
\]
This metric measures if the recommender successfully implements an explicit user intervention~\cite{chen2012critiquing,luo2020latentlinearcritiquing}. High values indicate stronger controllability and more reliable interactive personalization, since the recommendation list adapts predictably in response to user refinements or corrections.
For example, if an edit requests ``less explicit lyrics'' and \(6\) of the next \(10\) recommended items satisfy the updated hard constraints, then \(\mathrm{EditSuccess@10}=0.6\).
From the \textbf{creator/platform side}, it means fewer wasted turns and resources, and more efficient user guidance. Conversely, low edit success implies that the user must invest more interaction effort before the system behaves as intended, which weakens both \emph{short-term satisfaction} and the perceived value of conversational control.

Let $q_\tau^{(e)}$ denote the $\tau$-th recommendation turn following an explicit user edit $e$. For a declared profile-fidelity threshold $\gamma\in(0,1]$, we define the adaptation latency as:
\[
\mathrm{AdaptationLatency}(e;\gamma)
=
\min
\left\{
\tau\ge 1 :
\mathrm{ProfileAlignment@K}(q_\tau^{(e)},u)\ge\gamma
\right\},
\]
with value $+\infty$ (or a right-censored horizon) if the threshold is never reached.
This metric measures how quickly the recommender adapts to the updated preference state induced by the user edit, and therefore serves as the conversational-efficiency counterpart of profile fidelity~\cite{jannach2021crs,luo2020latentlinearcritiquing}. Lower values are preferable, since they indicate that the system aligns with the edited user state in fewer interaction turns.
For example, if the recommender requires three recommendation turns before the recommendation profile reaches the desired alignment threshold, then \(\mathrm{AdaptationLatency}=3\).
From the \textbf{listener side}, low latency reduces friction and supports more fluid interaction. 
From the \textbf{creator/platform side}, it means fewer wasted turns and more efficient user guidance. 

\begin{table}[t]
\scriptsize
\renewcommand{\arraystretch}{1.1}
\rowcolors{2}{dzrblue!12}{white}
\caption{Summary of universal success dimensions and setting-specific diagnostics for LLM-driven music recommendation. The first block lists the universal metric families. The second block lists diagnostic bundles for ICL, RAG, and reasoning-based settings; these do not replace the universal layer, but explain configuration-specific variation around it.}
\label{tab:final_table4}
\begin{tabularx}{\linewidth}{@{}L{2.5cm} L{4.6cm} X@{}}
\toprule
\textbf{Dimension / setting} & \textbf{Representative metrics} & \textbf{What it captures} \\
\midrule
\multicolumn{3}{l}{\textit{Universal success dimensions (Section~\ref{subsec:success_overview})}}\\
\textbf{G1. Query adherence and groundedness} & $\mathrm{EntityGroundedness@K}$, $\mathrm{QueryAdherence@K}$, $\mathrm{QueryAlignment@K}$, $\mathrm{ConstraintHitRate@K}$, $\mathrm{ExplanationGroundedness}$  \newline (paired complement: $\mathrm{HallucinationRate@K}$) & Checks if returned items are real, query-faithful, and explained consistently. Higher values are preferred for the success metrics, whereas HallucinationRate@\(K\) should decrease. \\
\textbf{G2. Discovery quality} & $\mathrm{LongTailShare@K}$, $\mathrm{ARP@K}$, $\mathrm{ILD@K}$, $\mathrm{CatalogCoverage@K}$ & Captures exploration through long-tail exposure, lower popularity concentration, within-list diversity, and broader catalog use. The desired direction is query-conditioned, but in exploratory settings LongTailShare, ILD, and coverage should increase, whereas ARP should decrease. \\
\textbf{G3. Incremental value of personalization} & $\Delta_{\mathrm{profile}}$, $\mathrm{InterUserOverlap@K}$, $\mathrm{UserNovelty@K}$ & Measures if profile conditioning adds utility beyond generic behavior, differentiates users, and surfaces items that are new relative to prior exposure. Higher \(\Delta_{\mathrm{profile}}\) and $\mathrm{UserNovelty@K}$ are desirable, whereas lower $\mathrm{InterUserOverlap@K}$ indicates stronger personalization. \\
\textbf{G4. Profile fidelity and controllability} & $\mathrm{ProfileAlignment@K}$, $\mathrm{EditSuccess@K}$, $\mathrm{AdaptationLatency}$ & Measures fidelity to the user representation and responsiveness to explicit edits. Higher alignment and edit success are desirable, while lower latency indicates better controllability. \\
\textbf{G5. Cultural and linguistic coverage} & $\mathrm{Share@K}$, $\mathrm{TargetDisparity@K}$, $\mathrm{WorstGroupGap@K}$ & Evaluates cultural fit at the list level and robustness across groups. Share@\(K\) should track the intended target composition, while target disparity and worst group gap should decrease. \\
\textbf{G6. Classical relevance} & $\mathrm{Precision@K}$, $\mathrm{Recall@K}$, $\mathrm{DCG@K}$, $\mathrm{nDCG@K}$ & Provides the classical relevance layer and remains useful as a compatibility baseline, but should be interpreted jointly with G1--G5 rather than on its own. \\
\midrule
\multicolumn{3}{l}{\textit{Setting-specific diagnostics (Section~\ref{subsec:setting-diagnostics})}}\\
\textbf{S-ICL (few-shot prompting)} & Shot calibration curves, PromptSensiScore, Rank-Biased Overlap, Preference injection tests, Instruction dominance over noisy/conflicting examples & Diagnoses if few-shot examples improve performance in a stable and controllable way, or instead introduce prompt brittleness and instruction drift. \\
\textbf{S-RAG (retrieval-augmented)} & EvidenceGrounding@\(k\), Doc-swap test: \(\Delta_{\mathrm{doc}}\), Recall@\(M\), TTFS (Time-to-First-Surface), Day-0 Lift for new releases, Token cost & Diagnoses if retrieved evidence is actually used, if retrieval supports freshness, and what efficiency costs are incurred. These diagnostics should be read together with the universal grounding metrics in G1. \\
\textbf{S-CoT / ToT (reasoning-based)} & Reasoning faithfulness, Utility gain over direct decoding \((\Delta U_{\mathrm{CoT}})\), Flow coherence, Self-consistency uplift & Diagnoses if explicit reasoning improves recommendation quality, remains faithful to catalog evidence, and supports coherent sequence construction when ordering matters. \\
\bottomrule
\end{tabularx}
\end{table}

\paragraph{G5. Cultural and linguistic coverage.}

For cultural and linguistic evaluation, we consider groupings that are relevant to the request or evaluation policy, such as language, region, country, or music community. Here, \emph{music community} refers to a socially, geographically, or stylistically situated group of musical practice, such as Detroit techno, UK grime, Brazilian forr\'o, or a local punk community. Let $\gamma(i)\in\mathcal{G}$ assign item $i$ to one such group under the chosen grouping scheme. For a group $g\in\mathcal{G}$, we define its observed share in the recommendation list as:
\[
\mathrm{Share@K}(g;q)
=
\frac{1}{K}
\sum_{j=1}^K
\mathbf{1}\!\left[\gamma(i_j)=g\right].
\]

This metric measures the proportion of the top-$K$ recommendations that belong to group $g$~\cite{singh2018fairness,zehlike2017fair}. It is a descriptive composition measure rather than a success criterion by itself: its interpretation depends on the target implied by the user request or by the evaluation policy. For example, for the request ``Brazilian forr\'o with accordion solos,'' a high share of Brazilian or Portuguese-language items would be expected. By contrast, for a request asking for a balanced mix across Latin American regions, the observed shares should be compared against the desired target distribution.

Next, let $\pi_g(q)$ denote the target group distribution implied by a user request or by a declared evaluation policy such as a quota.
We define disparity from the target composition as:
\[
\mathrm{TargetDisparity@K}(q)
=
\sum_{g\in\mathcal{G}}
\left|
\mathrm{Share@K}(g;q)-\pi_g(q)
\right|.
\]
This metric measures the discrepancy between the observed group composition of the recommendation list and the intended target distribution, following fairness-of-exposure and equity-of-attention formulations~\cite{singh2018fairness,biega2018equityattention}. Lower values indicate that the returned recommendations more closely match the desired cultural or linguistic balance.
For example, if a bilingual request asks for a balanced Spanish--Portuguese mix but the returned list is almost entirely Spanish, then the target disparity score will be high.
From the \textbf{listener side}, low disparity means that the recommendation list better matches the cultural or linguistic composition implied by the request. From the \textbf{creator/platform side}, it indicates less systematic underexposure of relevant groups. Conversely, high disparity suggests that the system may still appear relevant in a narrow sense while failing to provide cultural fit, identity recognition, or equitable exposure for underrepresented languages, regions, or scenes.

Let $Q_g\subseteq Q$ denote the evaluation slice associated with group $g\in\mathcal{G}$. We define the slice-level failure rate as:
\[
F_g
=
1-
\frac{1}{|Q_g|}
\sum_{q\in Q_g}
\mathrm{ConstraintHitRate@K}(q).
\]
We then define the worst-group gap as:
\[
\mathrm{WorstGroupGap@K}
=
\max_{g\in\mathcal{G}} F_g
-
\min_{g\in\mathcal{G}} F_g.
\]
This metric measures how unevenly recommendation failures are distributed across groups, and is thus closely related to worst-group robustness under distribution or group shift~\cite{sagawa2020groupdro,singh2018fairness}. Lower values indicate that no particular user group experiences systematically worse recommendation quality relative to others.
For example, if one language slice exhibits a failure rate of \(0.45\) while another exhibits a failure rate of \(0.10\), then the worst-group gap is \(0.35\).
From the \textbf{listener side}, low values suggest that minority or non-dominant users are not substantially underserved. From the \textbf{creator/platform side}, they indicate fairer cross-group performance and a more balanced distribution of exposure. Conversely, a large gap means that average performance may still look acceptable while at least one group experiences persistently weaker adherence, thereby weakening inclusion, identity fit, and long-term trust in the platform.

\begin{table}[t]
\centering
\rowcolors{2}{dzrblue!12}{white}
\caption{Schema of TalkPlayData--2 dataset.}
\label{tab:talkplaydata2-schema}
\small
\begin{tabularx}{\linewidth}{@{}l l X@{}}
\toprule
\textbf{Variable} & \textbf{Type} & \textbf{Description} \\
\midrule
\texttt{session\_id} & string &
A string identifier for the session (14–18 characters), typically combining the user ID and the date in the form \texttt{<userID>\_\_<date>} (e.g., \texttt{53726\_\_2011-12-26}). \\

\texttt{user\_id} & string &
A string identifier (2–6 digits) for the synthetic user participating in the session. This value can be used to relate multiple sessions involving the same user. \\

\texttt{session\_date} & string &
A date string (\texttt{YYYY-MM-DD}) indicating when the conversation took place.  Dates in the dataset span roughly from 2006-05-28 to 2018-12-31. \\

\texttt{user\_profile} & dict &
A dictionary describing demographic and preference attributes for the user. Example keys include \texttt{age}, \texttt{age\_group}, \texttt{country\_code}, \texttt{country\_name}, \texttt{gender}, \texttt{preferred\_language}, \texttt{preferred\_musical\_culture}, and \texttt{user\_split} (e.g., “train\_warm”). \\

\texttt{conversation\_goal} & dict &
A structured description of the listener’s objective. It contains a category code (A–K), a free-text description of the goal (\texttt{listener\_goal}), and a specificity label (LL, HL, LH or HH).  The categories correspond to discovery mechanisms such as audio-based, lyrical, mood-based, etc., and the specificity dimension indicates how precise the request and the target recommendation are. \\

\texttt{conversations} & list &
A list of dialog turns. Each element is a dictionary containing the \texttt{content} (either natural-language text or a track ID), the \texttt{role} (e.g., “user”, “music” or “assistant”), an optional \texttt{thought} field representing the agent’s reasoning, and the \texttt{turn\_number}.  This sequence captures the interaction history and the recommended tracks for the session. \\

\texttt{goal\_progress\_assessments} & list &
A list that assesses how each turn moves toward the conversation goal.  Each entry includes a \texttt{goal\_progress\_assessment} label (such as “MOVES\_TOWARD\_GOAL” or “DOES\_NOT\_MOVE\_TOWARD\_GOAL”) and the corresponding \texttt{turn\_number}.\\
\bottomrule
\end{tabularx}
\end{table}

\begin{table}[t]
\centering
\rowcolors{2}{dzrblue!12}{white}
\caption{Turn\textendash by\textendash turn evolution of the running example. The table distinguishes contextual turns from scored turns and makes explicit which request is being evaluated at each stage.}
\label{tab:running-example1-turns}
\small
\begin{tabularx}{\linewidth}{@{}c X L{5cm} X@{}}
\toprule
\textbf{Turn} & \textbf{User request summary} & \textbf{System response} & \textbf{Role in evaluation} \\
\midrule
1 &
Cover\textendash art search only: the user remembers a striking, unsettling rock album cover, but does not yet specify genre, era, or exclusion constraints. &
Talking Heads track (\texttt{1i6N76fftMZhijOzFQ5ZtL}); assistant explicitly recommends \emph{Psycho Killer}. &
Context only. This turn is not scored under the present G1 setup because the request is still too underspecified for strict constraint extraction. \\

2 &
The user clarifies that the goal is to identify an album from its unusual cover art, not to continue receiving Talking Heads recommendations. &
Talking Heads track (\texttt{38Ngied9rBORlAbLYNCl4k}); assistant recommends \emph{Once in a Lifetime}. &
Context only. This turn still lacks a stable, checkable metadata constraint set, so it is not included in the quantitative G1 average. \\

3 &
The user adds explicit metadata constraints: \emph{alternative or industrial rock} and \emph{late 1990s}; the request is also clearly for a \emph{different album}. &
Talking Heads track (\texttt{2VNfJpwdEQBLyXajaa6LWT}); assistant recommends \emph{Burning Down the House}. &
\textbf{Scored.} First formally evaluated turn. We score it with \(C(q_3)=\{c_{\text{era}},c_{\text{style}}\}\). \\

4 &
The user now makes the exclusion explicit: \emph{not Talking Heads}, while repeating \emph{alternative/industrial rock} and \emph{late 1990s}. &
Talking Heads track (\texttt{7CqleiaEqHVazV19P532X9}); system again returns a Talking Heads item. &
\textbf{Scored.} Second evaluated turn. We score it with \(C(q_4)=\{c_{\text{avoid}},c_{\text{era}},c_{\text{style}}\}\). \\

5 &
The user repeats the same constraints once more, emphasizing that the target is a provocative late\textendash 1990s alternative/industrial album with a red\textendash haired androgynous figure. &
Marilyn Manson track (\texttt{1KQxH1Z1BiSo3MMukVpRfl}); the system finally switches away from Talking Heads and returns \emph{The Dope Show}. &
\textbf{Scored.} Third evaluated turn. We score it with the same constraint set as turn 4, \(C(q_5)=\{c_{\text{avoid}},c_{\text{era}},c_{\text{style}}\}\). \\

6 &
The user explicitly confirms success: \emph{``That's the one! Marilyn Manson's `Mechanical Animals'!''} &
System continues with another track from the correct album neighborhood. &
Post\textendash success continuation. This turn confirms that the identification goal has been achieved, but it is not part of the present G1 scoring window. \\
\bottomrule
\end{tabularx}
\end{table}

\subsubsection{Running Examples}
The universal metrics are most informative when they are read as a coordinated diagnostic set rather than as isolated scores. The following two case studies instantiate the proposed framework on concrete conversations from TalkPlayData--2 \cite{doh2025talkplay}. The first example focuses on G1 and shows that catalog grounding does not imply request satisfaction. The second example focuses on G2 and G3 and shows that measurable personalization gain does not necessarily imply discovery. Table~\ref{tab:talkplaydata2-schema} summarizes the dataset fields used in the analysis, and Table~\ref{tab:running-example1-turns} summarizes the turn structure of the first session.

\paragraph{Running Example 1: Per–turn G1 evaluation on a TalkPlayData–2 session}
\label{subsec:running_ex1}
 Session \texttt{12286\_\_2016-02-17} involves a listener from Belarus who is trying to identify a specific album from a partial memory of its cover art. The conversational goal is thus not generic recommendation, but constraint refinement over time until one specific target can be recovered. This makes the session especially suitable for illustrating the distinction between \emph{groundedness} and \emph{query adherence}: the system may keep returning real catalog items while still failing to satisfy the actual request. An overview of the session is presented below.

\begin{tcolorbox}[colback=dzrorange!10!white,colframe=dzrorange!50!white,coltitle=black,
  title={Session \texttt{12286\_\_2016-02-17}: user, goal, and request evolution},
  fonttitle=\bfseries, breakable]
\small
\textbf{User profile.}  A 25‑year‑old male from Belarus (\texttt{country\_code = BY}) with English as preferred language and a stated interest in Western alternative rock.

\textbf{Conversation goal.}  Category C, high specificity: the listener wants to identify a single album based on its “distinctive and provocative cover art featuring an androgynous figure”.

\textbf{Why this session is useful.}  The user begins with a vague cover description and gradually introduces precise metadata constraints (genre, era, exclusion).  Meanwhile, the system repeatedly suggests Talking Heads until the final scored turn.  This illustrates how an LLM can be both grounded and non‑adherent at intermediate turns.

\textbf{Turn progression.}
\begin{itemize}
  \item \textbf{Turns 1–2 (context):} broad description and clarification that the task is album identification, not favourite‑artist recommendation.
  \item \textbf{Turn 3:} new constraints “alternative/industrial rock” and “late 1990s.”
  \item \textbf{Turn 4:} adds the exclusion “not Talking Heads.”
  \item \textbf{Turn 5:} repeats all constraints; the system finally recommends Marilyn Manson.
  \item \textbf{Turn 6:} user confirms \emph{Mechanical Animals}.
\end{itemize}
\textbf{Goal–progress annotations.}  The early turns are annotated as not moving toward the goal, which is consistent with the repeated Talking Heads recommendations and with the user’s growing frustration \cite{choi2025talkplaydata}.
\end{tcolorbox}

\textit{Metric computation.} In this example, we compute the G1 metrics only for turns in which the user request can be translated into explicit, checkable constraints. Turns~1--2 are therefore used as conversational context, but are not included in the quantitative metric computation. At that stage, the user only describes a remembered album cover, and the request does not yet specify stable catalog-level constraints such as genre, era, artist exclusion, or release type.

The first turn included in the metric computation is Turn~3. At this point, the user introduces two constraints that can be operationalized using catalog metadata: an era constraint and a style constraint. We thus define:
\[
C(q_3)=\{c_{\text{era}},c_{\text{style}}\},
\qquad
c_{\text{era}}:\text{year}\in[1996,1999],
\qquad
c_{\text{style}}:\text{alternative/industrial rock}.
\]

Subsequent turns are evaluated in the same way, using the updated constraint set implied by the latest user request. For instance, when the user explicitly indicates that the target is not Talking Heads, an additional exclusion constraint is added to the constraint set.

At Turn~4, the exclusion constraint becomes explicit:
\[
C(q_4)=\{c_{\text{avoid}},c_{\text{era}},c_{\text{style}}\},
\qquad
c_{\text{avoid}}:\text{artist}\neq \text{Talking Heads}.
\]
Turn~5 keeps the same scored constraint set:
\[
C(q_5)=\{c_{\text{avoid}},c_{\text{era}},c_{\text{style}}\}.
\]

\textit{Turn-level results.}
At Turns~3 and~4, the returned items resolve correctly to catalog entities, so:
\[
\mathrm{EntityGroundedness@1}(q_3)=1, \mathrm{EntityGroundedness@1}(q_4)=1
\]
\[
\mathrm{HallucinationRate@1}(q_3)=0, \mathrm{HallucinationRate@1}(q_4)=0
\]
However, the recommendations remain non-adherent:
\[
\mathrm{QueryAdherence@1}(q_3)= 0, \mathrm{QueryAlignment@1}(q_3)=0, \mathrm{ConstraintHitRate@1}(q_3)=0.
\]
\[
\mathrm{QueryAdherence@1}(q_4)=0, \mathrm{QueryAlignment@1}(q_4)=0, \mathrm{ConstraintHitRate@1}(q_4)=0.
\]
This means that the system is fully grounded and simultaneously fully wrong with respect to the operative request. If the explanation text names the same returned Talking Heads track, then \(\mathrm{ExplanationGroundedness}(q_3)=\mathrm{ExplanationGroundedness}(q_4)=1\), which is itself informative: explanation consistency does not rescue recommendation failure when the wrong entity is being explained faithfully.

At Turn~5, the system finally switches to the correct catalog neighbourhood and all G1 scores become maximal:
\[
\mathrm{EntityGroundedness@1}(q_5)=1,\qquad
\mathrm{HallucinationRate@1}(q_5)=0,
\]
\[
\mathrm{QueryAdherence@1}(q_5)=1, \mathrm{QueryAlignment@1}(q_5)=1, \mathrm{ConstraintHitRate@1}(q_5)=1.
\]

\textit{Session-level summary.}
Averaging over the scored turns \(t\in\{3,4,5\}\) gives
\[
\overline{\mathrm{EntityGroundedness@1}}=1,\qquad
\overline{\mathrm{HallucinationRate@1}}=0,
\]
\[
\overline{\mathrm{QueryAdherence@1}}=\frac{1}{3},\qquad
\overline{\mathrm{QueryAlignment@1}}=\frac{1}{3},\qquad
\overline{\mathrm{ConstraintHitRate@1}}=\frac{1}{3}.
\]

\textit{Interpretation.}
This example isolates the core rationale for reporting multiple G1 metrics rather than a single validity score. Catalog grounding would make the system appear strong, because every scored recommendation is executable and non-hallucinated. The adherence metrics reveal a different story: until the final turn, the model repeatedly fails to follow the actual request. The case thus show that \emph{short-term satisfaction} depends on both real-item grounding and faithful constraint following, and that the two should not be conflated.

\paragraph{Running Example 2: G2–G3 evaluation on an exploratory query}
\label{subsec:running_ex2}
Session \texttt{53726\_\_2011-12-26} involves a listener from Brazil who explicitly asks to discover \emph{new} artists with an intense or dramatic post-hardcore vibe. The conversation is thus discovery-oriented. This makes the session appropriate for separating G2 and G3: the recommender may still exploit prior taste information successfully while failing the user’s current novelty goal. The session overview is presented below.

\begin{tcolorbox}[colback=dzrorange!12!white,colframe=dzrorange!50!white,coltitle=black,
  title={Session \texttt{53726\_\_2011-12-26}: user, goal, and request evolution},
  fonttitle=\bfseries, breakable]
\small
\textbf{User profile.}  A 20‑year‑old female from Brazil whose listening history is dominated by Western alternative rock.

\textbf{Conversation goal.}  The listener states she wants to discover “new artists” with an intense or dramatic post‑hardcore vibe.

\textbf{Why this session is useful.}  This conversation highlights how novelty requests can be misinterpreted by a recommender.  The user begins by asking to \emph{“discover some new artists”} and specifies a desired mood (\emph{intense or dramatic}).  Because a previous session involved an Alesana track, the system continues to recommend Alesana songs: its first response notes that it is \emph{“sticking with”} that band because the listener enjoyed it.  Later, the user strengthens the novelty constraint (e.g., \emph{“by an artist I haven’t heard before”} or \emph{“not Alesana”}), yet the system again suggests Alesana.  This mismatch between user intent and system behavior makes the session a concrete example of a recommendation failure.

\textbf{Why this session is useful.}  The system repeatedly suggests Alesana, even after the user explicitly excludes that artist.  This exposes a divergence between personalization (reusing prior taste) and discovery (seeking novelty).

\textbf{Turn progression.}
\begin{itemize}
  \item \textbf{Turn 1:} user asks for new, intense music; the system plays Alesana.
  \item \textbf{Turn 2:} reiterates novelty and vibe; the system plays Alesana again.
  \item \textbf{Turn 3:} explicitly “not Alesana”; the system still returns Alesana.
  \item \textbf{Subsequent turns:} the user continues to seek new artists that fit the requested vibe; however, goal\,progress annotations for the early turns mark them as \emph{not moving toward the goal}, reflecting the recommender’s inability to adjust to the novelty constraint.

\end{itemize}
\textbf{Goal\textendash{}progress annotations.}  According to the dataset’s goal\,progress assessments, the first turn is unlabelled, the second turn moves toward the goal, but turns three and four are annotated as \emph{does\_not\_move\_toward\_goal}.  These annotations align with the user’s growing frustration at being recommended music from the same artist despite asking for new artists.

\end{tcolorbox}
\textit{Scoring setup.}
The scored turn is Turn~3, where the request can be operationalized as
\[
q_3=\text{``something by an artist I have not heard before, with intense post-hardcore energy''},
\]
with:
\[
C(q_3)=\{c_{\text{novel}},c_{\text{style}},c_{\text{vibe}}\},
\]
where,
\[
c_{\text{novel}}:\text{artist}\neq\text{Alesana},\qquad
c_{\text{style}}:\text{post-hardcore},\qquad
c_{\text{vibe}}:\text{intense/dramatic}.
\]

To illustrate how the list-level metrics are computed, suppose that the recommender returns the following top-3 list for query $q_3$: $R_K(q_3,u)=(i_1,i_2,i_3)$.
In this illustrative case, assume that all three recommended items are valid Alesana tracks. The example is not intended to reproduce the full hidden ranking of the system, but to make the metric calculations transparent for the observed failure mode.

\textit{Metric values.}

Because all three recommended items are valid catalog entries,
\[
\mathrm{EntityGroundedness@3}(q_3)=1,
\qquad
\mathrm{HallucinationRate@3}(q_3)=0.
\]

Let's assume that each returned track satisfies the style and vibe requirements but violates the novelty requirement, i.e., the artist is still Alesana. If novelty is treated as a hard constraint, then no item satisfies the full hard constraint set:
\[
\mathrm{QueryAdherence@3}(q_3)=0,
\qquad
\mathrm{ConstraintHitRate@3}(q_3)=0.
\]

By contrast, soft query alignment can still assign partial credit. With equal weights over the three requirements
\(\{c_{\mathrm{novel}},c_{\mathrm{style}},c_{\mathrm{vibe}}\}\), each item satisfies two out of three requirements, and thus:
\[
\mathrm{QueryAlignment@3}(q_3)=\frac{2}{3}.
\]

Thus, the list partially matches the musical description of the request, since it captures the desired style and vibe, but it fails the explicit novelty requirement. This illustrates why strict adherence and soft alignment should be reported separately.

From the G2 perspective, assume that all three Alesana tracks belong to the popularity head and are near-redundant under the selected dissimilarity function. Then:
\[
\mathrm{LongTailShare@3}=0,\qquad
\mathrm{ARP@3}=1,\qquad
\mathrm{ILD@3}=0.
\]
For a singleton evaluation set \(Q=\{q_3\}\), catalog coverage is:
\[
\mathrm{CatalogCoverage@3}(Q)=\frac{3}{|I|},
\]
which is formally correct, but not substantively informative on its own.
From the G3 perspective, let us define the utility $U$ of the items in the recommendation list as the mean satisfaction of the three facets \(\{c_{\text{novel}},c_{\text{style}},c_{\text{vibe}}\}\). Since each Alesana track satisfies two of the three, the full-system utility is:
\[
U(\mathrm{R}(q_3,u))=\frac{2}{3}.
\]
If a profile-ablated baseline satisfies none of the three facets, then
\[
U(\mathrm{R}(q_3,\emptyset))=0,\qquad
\Delta_{\mathrm{profile}}=\frac{2}{3}.
\]
At the same time, because the returned items belong to an already-known artist,
\[
\mathrm{UserNovelty@3}(q_3,u)=0.
\]
This means that the recommender extracts useful information from the user profile, but uses it to reinforce familiarity rather than support discovery. Even if InterUserOverlap were low against a very different user, that would not change the central point: \emph{inter-user differentiation} and \emph{within-user novelty} are distinct properties.

\textit{Interpretation.}
This case shows why G2 and G3 should not be collapsed into a single notion of recommendation quality. The system exhibits positive personalization gain, because conditioning on the user profile improves alignment with the desired post-hardcore style and mood. Yet it fails the discovery goal completely, because the returned list remains head-heavy, redundant, and non-novel for the current user. The example therefore illustrates a central tension in music recommendation: a system can personalize successfully in the sense of exploiting past taste, while still failing to create \emph{long-term value} through exploration.

\subsubsection{Setting-specific Diagnostics}\label{subsec:setting-diagnostics}

In the following, we discuss setting-specific diagnostic metric families for
\emph{few-shot in-context learning (ICL)}, \emph{retrieval-augmented generation (RAG)},
and \emph{chain-/tree-of-thought (CoT/ToT)}. By setting-specific diagnostics, we mean
metrics that are not universal to all LLM-based music recommenders, but are useful for
analyzing failure modes introduced by a particular configuration. For example, ICL
requires diagnostics for prompt sensitivity and example selection, RAG requires
diagnostics for evidence grounding and retrieval freshness, and CoT/ToT requires
diagnostics for reasoning faithfulness and sequence coherence. Table~\ref{tab:final_table4}
provides a concise summary and reference frame for these diagnostics in NL-based
evaluation of LLM-driven MRSs.

\paragraph{Few-shot In-Context Learning (S-ICL)} The evaluation of few-shot ICL might make use of several specialized metrics to assess the \textit{stability} and \textit{utility} of few-shot prompting in music recommendation. For instance, \textit{shot calibration}~\cite{zhang2024calibration} tests how performance changes as the number of examples (shots) varies; ideally, adding more examples should not drastically mislead the model.~\citet{zhang2024calibration} show that as we increase the number of in-context examples, models often become more miscalibrated before improving, with particularly poor calibration in low-shot regimes, highlighting the need to monitor this trade-off. The authors suggest measuring calibration curves across different shot counts (e.g., 0, 3, 5, 7) to ensure the model benefits from additional examples. %predictably. 
Zhou et al.~\cite{zhou2024batch} propose batch calibration to address the contextual bias that arises from varying numbers of in-context examples, showing that proper calibration techniques can recover performance degradation caused by prompt brittleness across different shot configurations.

\textit{Stability to prompt perturbations} is another critical dimension: the model’s recommendations should remain consistent under minor prompt modifications (e.g., reordering or paraphrasing the examples). For instance, Zhuo et al.~\cite{zhuo2024prosa} introduce PromptSensiScore (PSS), which quantifies prompt sensitivity as the average discrepancy in LLM responses when confronted with semantic variants of the same instruction. 
Their findings reveal that few-shot examples can alleviate sensitivity, with the most pronounced reduction occurring in the transition from zero-shot to one-shot settings. For measuring list-based recommendation stability, rank-biased overlap~\cite{webber2010rbo} provides a complementary assessment of ranking consistency under perturbations.

Additionally, \textit{preference sensitivity} could be evaluated to ensure that when user preference cues (likes/dislikes) are given as examples, the recommender’s output shifts appropriately along the intended facets (such as recommending more of a liked genre and fewer disliked styles). For instance, if a user’s example indicates love of jazz and dislike of loud rock, the subsequent recommendations should reflect those biases measurably. 
Yu et al.~\cite{yu2024icpl} propose In-Context Preference Learning (ICPL), showing that LLMs can learn reward functions from iteratively placed human feedback in context, achieving orders of magnitude greater sample efficiency than traditional preference-based methods. Their work shows that properly designed ICL can capture preference signals and translate them into consistent recommendation adjustments, validating that LLMs possess native preference-learning capabilities that enable few-shot adaptation to user tastes.

Finally, \textit{instruction–example alignment} might check that the system continues to obey the explicit request instructions even when some in-context examples might be ambiguous or in conflict with those instructions. In other words, the user’s prompt requirements (e.g., “no mainstream hits”) must dominate over any contradicting signals from the examples (e.g., in case these include predominantly chart music). 
Zhou et al.~\cite{zhou2024batch} show that LLM outputs can be sensitive to seemingly minor prompt-design choices, including formatting, label wording in classification-style prompts, and the selection of in-context examples. Such factors can cause the model to deviate from the intended instruction. Zhuo et al.~\cite{zhuo2024prosa} further show that instruction-following fidelity is related to model confidence, with higher-confidence generations being more robust to prompt variations. In the context of ICL-based music recommendation, these findings motivate robustness checks that vary prompt templates, example order, and example selection.

\setcounter{enumi}{6}

\paragraph{Retrieval-Augmented Generation (S-RAG)}

The evaluation of an S-RAG pipeline spans multiple aspects across both the retrieval and recommendation stages. As noted by \citet{es2024ragas}, evaluating RAG systems requires assessing the retrieval module’s ability to fetch relevant context and the generation module’s ability to use that evidence. For the retrieval stage, one can use standard IR metrics (e.g., Recall@$K$ of known relevant items) to gauge if the LLM is retrieving useful evidence. At the recommendation generation stage, the primary focus is on \textit{grounding}: conditioning the LLM on up-to-date catalog facts should reduce hallucinations and ensure that recommended tracks exist and meet the user’s criteria. To quantify groundedness, one might define \emph{EvidenceGrounding@$K$} as the proportion of the top-$K$ recommended items whose key attributes can be justified by the retrieved items.
This metric is inspired by faithfulness and citation-based evaluations proposed for RAG outputs~\cite{es2024ragas,asai2024self}.  

Another diagnostic is a document-swap test, denoted $\Delta\textit{doc}$, where we perturb the retrieved context (e.g., replacing a percentage of relevant items with irrelevant ones) to see if the model’s recommendations change. 
A strong drop in recommendation quality after such a swap would indicate that the model truly relies on the retrieved evidence, whereas an invariant output would suggest that it ignores the provided items.
This methodology is analogous to input-perturbation analyses used to study how LLMs utilize long contexts~\cite{liu2024lost}. 

Beyond grounding, we also need to evaluate the \emph{freshness}, \emph{breadth}, and \emph{efficiency} of RAG-based recommendations. For timeliness, {inspired by} work on time-aware evaluation and recency effects in RSs~\citep{campos2014time,deldjoo2024understanding}, one might introduce freshness and temporal-bias metrics such as \emph{Time-To-First-Surface (TTFS)} and \emph{Day-0 Lift}, whose aim is to summarize how quickly new releases first appear in either the retrieved pool or the final top-$K$ list, and how much extra exposure they receive on their release day. For example, a low TTFS and high Day-0 Lift would indicate that a newly released track already shows up quickly and prominently in users’ recommendations instead of only weeks later.

 To understand which parts of the catalog the LLM actually uses, we monitor simple coverage-style metrics {inspired by} beyond-accuracy evaluation~\citep{shen2022evaluation,kaminskas2016diversity}, notably metrics such as \emph{UniqueArtists@$K$} counting distinct artists in the top-$K$ recommendations, or \emph{EntitiesResolvedShare} measuring the fraction of recommended items that can be matched to real catalog entities (i.e., non-hallucinated tracks). 
 
 Finally, it is important to monitor the inference latency $L$ (end-to-end response time), the total token budget $B$ (number of tokens processed per query, including prompt, retrieved evidence, and generated output), and the evidence token ratio $B_D/B$ (share of the budget devoted specifically to retrieved documents). An effective S-RAG configuration should improve freshness and catalog coverage at both retrieval and recommendation stages, without incurring prohibitive cost, while still satisfying core recommendation goals such as accuracy, novelty, and personalization.

\paragraph{Chain-of-Thought Reasoning (S-COT)}
Complex reasoning for playlist curation requires both verifiable reasoning steps and coherent output. Here, one might use \emph{Reasoning Faithfulness}, which measures intermediate steps verifiable against catalogs \cite{paul2024making}. In contrast, \emph{Reasoning Impact} ($\Delta U_{\text{CoT}}$) can be used to quantify the utility gain from explicit reasoning chains compared to direct answers \cite{wei2022chain}. \textit{Self-consistency Uplift} evaluates improvements from aggregating outputs across multiple reasoning paths \cite{wang2022selfconsistency}. Overall, one can state that the reasoning metrics integrate with query fidelity, diversity, and controllability dimensions.

\subsubsection{Measures of Risk}
Beyond success-oriented metrics, which capture how well a system fulfills user intent and provides relevant, grounded recommendations, it is equally important to account for potential \emph{risks} and undesirable behaviors arising from generative models in music recommendation. 
For this reason, we include a complementary set of \textbf{diagnostic metrics and dimensions} addressing issues such as hallucination, exposure skew, privacy and stereotyping, automatic-judge bias, and cross-lingual failure. 
These diagnostics provide a broader understanding of system robustness, fairness, and trustworthiness—dimensions that classical success metrics alone cannot capture.
We identify the following risks:

\highlightbox{ 

\begin{description}
\item[\textbf{R1. Hallucinations and entity errors.}] This dimension measures instances of fabrication (generating non-existent entities), misattribution (assigning incorrect properties to real entities), or failed resolution against the catalog or an external knowledge base.

\item[\textbf{R2. Popularity, temporal, and language bias.}] This dimension evaluates systematic disparities in exposure across stratified groups, such as mainstream vs. long-tail content or Western vs. non-Western music.

\item[\textbf{R3. Profile hazards.}] This dimension assesses the system's sensitivity to variations in user profile representation, including robustness to profile edits (addition/removal of preference signals), degradation with truncated profiles (measuring utility loss when profile length is constrained), and the impact of noisy or contradictory preference signals. 

\item[\textbf{R4. Evaluator bias.}] When LLMs serve as automatic evaluators (LLM-as-judge), this dimension quantifies systematic divergence between LLM-based assessments and human ground-truth judgments.

\item[\textbf{R5. Sampling-bias amplification.}] This dimension evaluates how the composition of in-context examples systematically affects exposure distribution for specific groups (e.g., language, genre, or artist popularity tier). 

\item[\textbf{R6. Order brittleness and query contradiction.}] These criteria quantify recommendation instability arising from perturbations in example ordering or conflicts between explicit instructions and implicit example patterns. 

\item[\textbf{R7. Privacy leakage and reproducibility.}] These complementary dimensions assess seed example memorization in outputs (privacy risk) and generation determinism (reproducibility).  

\item[\textbf{R8. Implicit-feedback noise.}] This dimension quantifies how noisy implicit signals, e.g., skips or incomplete plays, in seed examples affect recommendation utility. 
\end{description}
}

\paragraph{R1. Hallucinations and entity errors.} 
Ji et al.~\cite{ji2023hallucination} provide a comprehensive taxonomy distinguishing between factuality hallucinations (contradicting established knowledge) and faithfulness hallucinations (inconsistency with provided context). For RAG systems, counterfactual robustness testing can be applied: evaluating if the model appropriately handles retrieved documents containing incorrect or contradictory information, or measuring if outputs change when relevant documents are replaced with irrelevant ones~\cite{es2024ragas}. 
High-quality RAG systems should show substantial sensitivity to document relevance, i.e., changing outputs significantly when retrieved context changes while maintaining the ability to reject answers when retrieved documents lack relevant information. In MRSs, this can be quantified as the proportion of recommended tracks that do not exist in the catalog or have incorrectly attributed metadata such as artist, album, or genre.

\paragraph{R2. Popularity, temporal, and language bias.} Abdollahpouri et al.~\cite{abdollahpouri2019managing} introduce the Average Recommendation Popularity (ARP) metric: $\text{ARP} = \frac{1}{|U|}\sum_{u \in U}\frac{\sum_{i \in R_u}\phi(i)}{|R_u|}$, where $\phi(i)$ represents item $i$'s popularity score and $R_u$ is the recommendation list for user $u$. They show that collaborative filtering algorithms exhibit systematic bias toward popular items, with ARP values significantly exceeding catalog median popularity. LLMs may exacerbate this bias or introduce new types of bias. Deldjoo et al.~\cite{deldjoo2024understanding} show that ChatGPT-based RSs exhibit provider fairness issues, temporal instability, and recency bias in music and movie recommendations, evaluating consistency through repeated queries and temporal drift analysis. Sguerra et al.~\cite{sguerra2025biases} reveal that LLM-generated musical taste profiles exhibit systematic quality variations across user mainstreamness and cultural diversity, measuring profile accuracy through genre precision and user study evaluations. 

\paragraph{R3. Profile hazards.}  Wang et al.~\cite{wang2025diagnostic} introduce profile defect diagnostics, categorizing issues as inaccurate (contradictory preferences), incomplete (missing critical attributes), or both. Evaluation involves measuring recommendation stability (via Rank-Biased Overlap~\cite{webber2010rbo}) across profile perturbations and quantifying the accuracy-length trade-off: plotting recommendation quality (nDCG or Recall) against profile token length to identify optimal compression points. For LLM-based systems, this includes assessing if prompt length constraints (e.g., 2048 vs. 8192 tokens) significantly degrade personalization quality.

\paragraph{R4. Evaluator bias.} Zheng et al.~\cite{zheng2023judging} show that LLM judges exhibit multiple forms of bias: position bias (where GPT-4 shows only 65\% consistency when answer positions are swapped, favoring the first position in 30\% of cases), verbosity bias (where judges may prefer longer responses, with GPT-4 showing 8.7\% susceptibility to verbosity attacks compared to over 90\% for weaker models), and self-enhancement bias (where GPT-4 exhibits approximately 10\% higher win rates for its own outputs compared to human judgments). To measure evaluator bias, Zheng et al.~propose conducting position swap tests to assess consistency and comparing agreement rates between LLM-human pairs versus human-human pairs, finding that GPT-4 achieves over 80\% agreement with human preferences when biases are mitigated. Additionally, Wang et al.~\cite{Huang2025empirical} provide comprehensive analysis of LLM-as-judge reliability across diverse evaluation scenarios. 

\paragraph{R5. Sampling-bias amplification.} Lin et al.~\cite{lin2024amplification} show that recommendation models memorize item popularity in the principal spectrum and amplify this bias. For ICL contexts, bias amplification can be quantified by comparing group representation in recommendations versus catalog baselines, measured as relative popularity lift: $\Delta \text{Pop}(g) = \frac{\text{Pop}_{\text{rec}}(g) - \text{Pop}_{\text{catalog}}(g)}{\text{Pop}_{\text{catalog}}(g)}$, where positive values indicate over-representation. 
Similarly, user-level popularity calibration~\cite{DBLP:conf/recsys/LesotaBWMLRS22} identifies popularity mismatches between historic user interactions $H_u$ and recommended items $R_u$, measured as Jensen-Shannon Divergence (JSD) between the popularity distribution of items in $H_u$ and $R_u$ (over a fixed number of popularity bins).
Evaluation should stratify across demographic dimensions and employ counterfactual seed sets to isolate causal effects of example composition.

\paragraph{R6. Order brittleness and query contradiction.} PromptSensiScore~\cite{zhuo2024prosa} measures sensitivity to prompt variants, finding that few-shot examples can alleviate instability with the most pronounced reduction in zero-to-one-shot transitions. Order brittleness can be measured via rank-biased overlap~\cite{webber2010rbo} across different orderings: $\text{OB} = 1 - \text{RBO}(R_{\pi_1}, R_{\pi_2})$, where $\pi_1, \pi_2$ represent different orderings of identical examples. Query contradiction is assessed by providing examples that violate stated constraints and measuring constraint adherence. When substantial instability is detected, fallback strategies include zero-shot prompting or explicit alignment scaffolds emphasizing instruction priority over example patterns.

\paragraph{R7. Privacy leakage and reproducibility.} Carlini et al.~\cite{carlini2021extracting} show that LLMs memorize training data, which can even be extracted in terms of verbatim sequences, particularly for repeated or distinctive content. In ICL contexts, leakage is measured as the proportion of seed example text surfacing in generated rationales or recommendations. For reproducibility, seed sets should be cryptographically hashed and logged alongside model version, temperature (with $T=0$ ensuring determinism), and sampling parameters. Reproducibility enables systematic auditing when recommendation quality degrades, verified by regenerating recommendations with identical inputs and measuring output consistency.  

\paragraph{R8. Implicit-feedback noise.} Hu et al.~\cite{hu2008collaborative} show that treating all implicit feedback uniformly degrades collaborative filtering performance, as signals like skips carry inherent ambiguity (dislike versus context inappropriateness). For ICL evaluation, one might compare recommendation quality across seed configurations: all interactions, positive-only signals, and confidence-weighted signals (e.g., high play completion rates), measuring utility via ranking metrics (nDCG@K) and diversity across configurations. When substantial quality degradation occurs with noisy seeds, filtering strategies should weight implicit signals by confidence indicators or exclude ambiguous interactions from ICL contexts.

\subsubsection{Further Considerations on User and Item Modeling} Since the goal of automatically generated user profiles or item descriptions is to replace traditional user or item representation in a recommendation setting, the most straightaway methods of evaluating their success is through traditional recommendation metrics such as Recall, and NDCG.
In production, if users or content editors are allowed to edit user profiles or music content descriptors, respectively, then the amount of correction required serves as a good indicator of the approach’s ability to accurately represent users or items.
However, most of the work on using LLMs to summarize users’ preferences in a scrutable way has focused on domains such as books, movies, or e-commerce~\cite{ramos2024transparent,  gao2024end, zhou2024language}, whereas music presents its own unique challenges.

Given the context-length limitations of LLMs, it becomes crucial to devise effective sampling strategies for selecting items that best capture users’ interests.
Moreover, in the absence of explicit textual feedback, user or item modeling with LLMs must rely on item metadata, whose quality can vary substantially~\cite{matrosova2024recommender}.
Also, various kinds of biases are present. In~\cite{sguerra2025biases}, authors show that profile quality, split into two dimensions (1) self-identification and (2) downstream recommendation, is biased towards certain types of music genres.
For example, the tested LLMs generated better descriptions of taste when the users consumed metal music that tends to be of American origin and is largely discussed on internet forums and magazines. More recent music tracks and content such as french Rap tended to be less well attended, indicating that due to the corpora used to train LLMs, some content types end up being underrepresented. These differences raise fairness questions, e.g., if some users do not perceive their tastes accurately reflected by the system,
they are less likely to trust the system.
In addition, some items might be less reliably described by an LLM.

Moreover, due to the dynamic nature of conversations in LLMs, established datasets for offline evaluation are generally insufficient. Instead, users' explicit indication of their queries, emotions, intents, etc.~need to be collected.
Finally, assessing the contribution of advanced user models that integrate personas or psychological factors with music preferences in  recommendation is considerably challenging, as disentangling the underlying factors is a non-trivial task.

\paragraph{LLMs and Long-Tail Items} 
LLMs struggle with \textit{long-tail entities}~\cite{huang2025prompting,hachmeier2025benchmark}, a challenge that is particularly present in the music industry, where a large portion of the catalog consists of such lesser-known music artists or tracks.
Recent research~\cite{huang2025prompting} has explored ways to address this limitation by introducing non-parametric knowledge into LLMs—either by augmenting prompts with relevant external information about the items or by injecting structured triplets from knowledge bases.
The authors' findings show that these approaches can significantly improve performance.
In particular, prompting LLMs with knowledge base triples tends to work better than just injecting external information about the items in the prompt, while combining both methods can substantially reduce hallucinations in the generated content.
{However, when long-tail items correspond to new tracks or artists for which such external information is not yet available, either textually or within knowledge bases, this strategy remains infeasible, and alternative approaches still need to be explored.
Evaluating performance on long-tail items is thus crucial, and at the very least, their impact should be systematically monitored.}

\paragraph{LLMs and Personas}
Personalization lies at the core of RSs.
In the context of LLMs, it could refer to their ability to represent different users or user groups, i.e., to go beyond an "average" listener or music and adapt to diverse music preferences and behaviors.
In other words, it reflects the model's steerability: its capacity to adjust its outputs based on the user.
In practice, steerability can be achieved through prompting, fine-tuning, or activation steering \cite{miehling2025evaluating}.
Prompting remains the most common approach, as fine-tuning and activation methods often require access to model internals or significant computational resources.
{However, \citet{miehling2025evaluating} show that prompt-based steerability remains highly challenging: 
it depends on model size, with larger models generally showing better steerability, and, more importantly, each model tends to be more responsive to specific subsets of persona dimensions by default.
Given these findings, how can we fairly assess an LLM's ability to adapt while keeping in mind their inherent bias toward certain dimensions?
Future research should measure not only overall quality but also how consistently models reflect individuals or groups, when used either in solutions or as evaluators.}